\documentclass[twocolumn]{aastex62}

\usepackage{threeparttable}
\usepackage{graphicx}
\usepackage{amsmath}

\newcommand{\mkms}{~\mathrm{km~s^{-1}}}

\newcommand{\mgtwo}{Mg~\textsc{ii}}
\newcommand{\fetwo}{Fe~\textsc{ii}}
\newcommand{\otwo}{O~\textsc{ii}}
\newcommand{\fetwostar}{Fe~\textsc{ii}$^*$}
\newcommand{\naone}{Na~\textsc{i}}
\newcommand{\router}{$r_{\rm outer}$}
\newcommand{\tkrs}{TKRS4389}
\newcommand{\seeing}{1.63''}
\newcommand{\lam}{$\lambda$}

\newcommand{\surfbrightlim}{$4.8 \times 10^{-19}$ erg s$^{-1}$ cm$^{-2}$ arcsec$^{-2}$}
\newcommand{\sbunits}{erg s$^{-1}$ cm$^{-2}$ arcsec$^{-2}$}
\defcitealias{Prochaska11_RT}{P11}

\shorttitle{\ion{Mg}{2} Emission from the CGM}
\shortauthors{Burchett et al.}

\begin{document}

\title{Circumgalactic \ion{Mg}{2} Emission from an Isotropic Starburst Galaxy Outflow Mapped by KCWI}

\correspondingauthor{Joseph N. Burchett}
\email{burchett@ucolick.org}

\author{Joseph N. Burchett}
\affiliation{University of California Observatories - Lick Observatory, University of California, Santa Cruz, CA 95064, USA}


\author{Kate H. R. Rubin}
\affiliation{San Diego State University, Department of Astronomy,
  San Diego, CA 92182, USA; Center for Astrophysics and Space Sciences,
  Department of Physics, University of California, San Diego, 9500 Gilman Drive, La Jolla, CA 92093, USA} 

\author{J. Xavier Prochaska}
\affiliation{University of California Observatories - Lick Observatory, University of California, Santa Cruz, CA 95064, USA; Kavli Institute for the Physics and Mathematics of the Universe, 5-1-5 Kashiwanoha, Kashiwa, 277-8583, Japan}, 

\author{Alison L. Coil}
\affiliation{Center for Astrophysics and Space Sciences,
  Department of Physics, University of California, San Diego, 9500 Gilman Drive, La Jolla, CA 92093, USA}

\author{Ryan Rickards Vaught}
\affiliation{Center for Astrophysics and Space Sciences,
  Department of Physics, University of California, San Diego, 9500 Gilman Drive, La Jolla, CA 92093, USA}

\author{Joseph F. Hennawi}
\affiliation{Department of Physics, Broida Hall, University of California at Santa Barbara, Santa Barbara, CA 93106, USA}

\begin{abstract}

We present spatially-resolved spectroscopy from the Keck Cosmic Web Imager (KCWI) of a star-forming galaxy at $z=0.6942$, which shows emission from the \mgtwo\ 2796, 2803 \AA\ doublet in the  circumgalactic medium (CGM) extending $\sim37$ kpc at 3$\sigma$ significance in individual spaxels (1$\sigma$ detection limit \surfbrightlim).  After deconvolution with the seeing, we obtain $5\sigma$ detections extending for $\sim31$ kpc measured in 7-spaxel (1.1 arcsec$^2$) apertures. Spaxels covering the galaxy stellar regions show clear P-Cygni-like emission/absorption profiles with the blueshifted absorption extending to relative velocities of $v = -800$ km s$^{-1}$; however, the P-Cygni profiles give way to pure emission at large radii from the central galaxy. We have performed three-dimensional radiative transfer modeling to infer the geometry and velocity and density profiles of the outflowing gas.  Our observations are most consistent with an isotropic outflow rather than biconical wind models with half-opening angles $\phi \leq 80^\circ$.  Furthermore, our modeling suggests that a wind velocity profile that decreases with radius is necessary to reproduce the velocity widths and strengths of \mgtwo\ line emission profiles at large circumgalactic radii.  The extent of the \mgtwo\ emission we measure directly is further corroborated by our modeling, where we rule out outflow models with extent $<30$ kpc.  

\end{abstract}



\section{Introduction} \label{sec:intro}

Simply stated, galaxy formation as we know it does not work without galactic winds.  These winds are now recognized as a fundamental activity of nearly all star-forming galaxies, a generalization following from a slew of spectroscopic surveys \cite[e.g.,][]{Pettini2001,Rupke2005a,Martin2005,Rubin2010,Rubin2014,Coil2011winds}, which span the majority of cosmic time and a wide diversity of galaxy properties.  Winds are revealed in these down-the-barrel galaxy spectra by the telltale signature of a systematic blueshift of gas relative to the nebular emission or stellar absorption features \citep{Heckman2000}.   The offset velocities range from tens to several hundreds of $\mkms$ and correlate weakly with galaxy mass and star formation rate \citep{Martin2005,Rupke2005b,Tremonti2007,Weiner2009,Rubin2014,Chisholm2016}.  The ubiquity and strength of galactic winds coupled with signatures of gas infall strongly suggest that such flows play an important role in regulating galaxy growth \citep{Steidel2003,Rubin2012,Martin2012,Rubin2014}.


Galactic winds have also become an essential ingredient in theoretical models of galaxy formation.  Initially, winds driven by star-formation feedback (e.g.\ supernovae) were introduced to maintain a turbulent interstellar medium \citep{CowieMckeeOstriker:1981}.  Now, winds are invoked as feedback mechanisms that regulate star formation and give rise to the observed distribution of stellar mass in galaxies.  Furthermore, these winds enrich the intergalactic medium (IGM) and distribute heavy elements throughout the Universe \citep{Oppenheimer:2006uq}.  While winds are a critical component in theoretical treatments, even the most sophisticated and/or idealized simulations of outflows \citep{Fielding2017,Schneider:2018_CGOLS} fail to capture all of the salient astrophysics, especially across the $\sim 10$\,Gyr duration of galaxy formation.\footnote{For an analytical treatment, also see \citet{Murray2011}.}  Among the many challenges these models face is a somewhat existential one: the body of galactic wind literature primarily comprises observations of \textit{cool gas} tracers.  Thus, the outflow-driving mechanism(s) must produce the observed high outflow velocities while not injecting so much thermal energy as to heat and further ionize the gas beyond the neutral or low-ionization states of the tracers (e.g., \naone, \mgtwo, and \fetwo).  Therefore, these theories demand empirical constraints to bound model parameter space and inform the wind prescriptions. Present-day scaling laws \citep[e.g.,\ the mass-metallicity relation;][]{Tremonti2004} and the distribution of galaxy properties \citep[e.g.,\ luminosity or stellar mass;][]{Willmer2006} offer useful targets for tuning wind parameters; however, more direct constraints on the physical properties of galactic winds are in desperate need.

To date, the majority of insight into the nature of galactic winds has been derived from absorption line analysis.  Early efforts focused on the gas kinematics, e.g.,\ estimating outflow speeds, and their scaling with galaxy properties \citep[e.g.][]{Rupke2005a,Martin2005,Weiner2005,Rubin2010}. Models were then introduced to match the flux profiles with velocity and thereby infer aspects of the density and velocity field \citep[e.g.,][]{MartinBouche2009,Steidel2010}. These models were necessarily simplistic and subject to significant uncertainties owing to various aspects of the experiment, such as (i) the down-the-barrel geometry yielding poorly constrained distances of the gas from the galaxy and (ii) accessing only a small set of ions and transitions.  Estimates for the physical extent of the wind and its mass were limited to order(s)-of-magnitude uncertainty.  

In the past several years, with the advent of {\it HST}/COS, new datasets spanning a wider range of ions through coverage of far-UV transitions at high spectral resolution have emerged  \citep{Heckman2015,Chisholm2016,Chisholm2017}. These provide better estimates of physical conditions of the gas, including its ionization state.  Adopting the reasonable ansatz that the gas is primarily ionized by the photon flux of the young stellar population driving the outflow, one may then estimate the distance to the medium.  
Assuming power-law density and velocity laws for the gas, these analyses suggest that the densities and covering fractions of the absorbing material decrease steeply with radius (i.e., $C_f \propto r^{-0.9}$ and $n \propto r^{-5.3}$), and that the launch radius of the flow occurs at distances $r \lesssim 100$ pc from the star-forming regions \citep{Chisholm2016,Chisholm2017}.  These constraints in turn imply that the rate of outward mass flow beyond several hundred parsecs drops to near zero \citep{Chisholm2016}, and hence that these winds do not result in significant enrichment of halo gas (with the caveat that they may undergo a change in phase on larger scales).


An alternative and more direct approach to assessing the radial density profile of winds is to image it in emission.  
Due to the overall low density of the material, the vast majority of studies pursuing this measurement have targeted nearby starbursting systems in narrow-band imaging or IFU surveys of 
 collisionally-excited transitions in the optical (e.g., [\ion{O}{3}], H$\alpha$;   
\citealt{HeckmanArmusMiley:1987,Lehnert:1999_M82,Veilleux2003,Robitaille2007,SharpBland-Hawthorn2010,RupkeVeilleux2013,Yoshida:2016_giantHaNebula,Leslie2017,McKinley:2018_CenA}). This work reveals that line emission extending over ${\sim}10-20$ kpc distances is common around these systems.  Millimeter and submm interferometry has likewise been useful for tracing the spatial extent of the cold component of nearby starburst outflows \citep[e.g.,][]{Walter2002,Bolatto2013,Geach2014,Leroy2015}.  
Spatially-resolved study of these transitions, however, has not been possible in star-forming systems beyond the very nearby universe.

In principle, winds may also be illuminated by  scattering of photons through resonant line transitions.    
Resonant absorption from Mg$^+$ and Fe$^+$ ions are commonly detected in absorption in galactic winds at rest-frame wavelengths $\lambda\lambda 2796, 2803$ and $\lambda \lambda 2586, 2600$, as these transitions are easily observable at $z>0.2$ with blue-sensitive ground-based instruments \citep[e.g.,][]{Rubin2010,Kornei2012,Erb2012}.  Re-emission of these absorbed photons (originally generated by the stellar continuum and/or line emission from \ion{H}{2} regions; \citealt{Henry2018})
therefore tracks the spatial extent of the wind material.  
For a given shell of gas with density $n$, width $\Delta r$, and velocity  gradient $\Delta v$, the optical depth scales as $\tau \sim n (\Delta v / \Delta r)^{-1}$ \citep{Sobolev1960}. The surface brightness profile of the emission is a complex radiation transfer problem, modulated by gas within the ISM and by dust throughout the environment (\citealt{Prochaska11_RT}; hereafter \citetalias{Prochaska11_RT}; \citealt{ScarlataPanagia2015,Carr2018}). Scattering analysis then provides unique insight on the distribution and  velocity of the outflow through assessment of the emergent emission profile.

\begin{figure}
\includegraphics[angle=0,width=\columnwidth,trim=150 100 150 100,clip=]{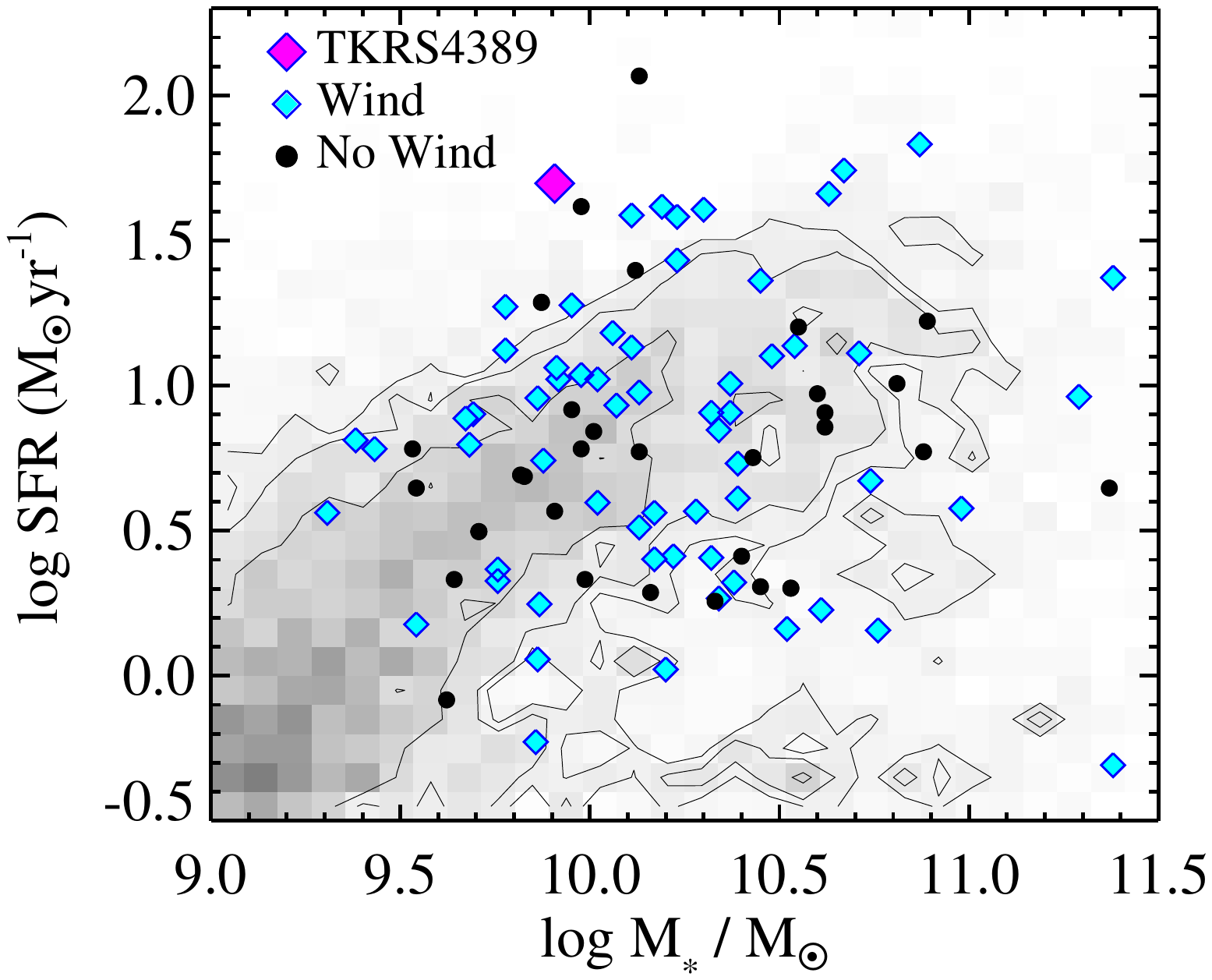}
\caption{Gray contours show the $\mathrm{SFR}-M_*$ distribution of the galaxy population at $0.4 < z < 0.8$ included in the \citet{Barro2011} catalog of multiwavelength photometry for galaxies in the Extended Groth Strip. The symbols show the galaxy sample studied in \citet{Rubin2014}: objects with outflows detected in \mgtwo\ or \fetwo\ absorption are marked in cyan, and objects without detected winds are marked in black. The location of TKRS4389 is marked with a large magenta diamond.  It lies well above the star-forming sequence at $z\sim0.7$. \label{fig.sfr_Mstar}}
\end{figure}

The observations presented here targeted \tkrs, a $z=0.6942$ star-forming galaxy originally identified by the Team Keck Treasury Redshift Survey \citep[TKRS;][]{Wirth2004}. \tkrs\ has a stellar mass log ($M_*/M_\odot$) = 9.9, star formation rate SFR = 50 $M_\odot$ yr$^{-1}$, a morphology indicative of a merger, and indications of an active galactic nucleus (Figure~\ref{fig.sfr_Mstar}; \citealt{Rubin:2010_GPG,Rubin2014}). In a previous study, we obtained slit spectroscopy with Keck/LRIS of \tkrs\ to leverage its bright $B$-band flux to probe the CGM of a close transverse foreground galaxy TKRS4259 ($z=0.4729$) in absorption \citep{Rubin:2010_GPG}. Analysis of the \tkrs\ spectral features revealed P-Cygni line emission in the \ion{Mg}{2}$\lambda\lambda 2796, 2803$ transitions and an outflow speed $\delta v \sim 800 \mkms$.  
In \citet{Rubin2011}, we demonstrated that this emission is spatially extended along the slit (oriented along the galaxy's minor axis) to a distance $\approx 7$\,kpc.  This was the first detection of spatially-extended \mgtwo\ emission from a star-forming galaxy, and offered the first direct measurement of the minimum spatial extent of a wind beyond the local universe.  
Our constraints were nevertheless limited to the emission located along the single slit position angle, and by the relatively low S/N of the spatially-resolved emission profile.

With the commissioning of the Keck Cosmic Web Imager \citep[KCWI;][]{KCWI}, a blue-sensitive integral field unit (IFU) spectrograph on the Keck~II telescope, we have returned to \tkrs\ 
to map the spatial extent and kinematics of this \ion{Mg}{2} emission across the sky. In the same data cube, we may also examine line emission from resonant and non-resonant \ion{Fe}{2} transitions, with the latter expected to be confined to the galaxy (\citealt{Rubin2011}; \citetalias{Prochaska11_RT}; but see \citealt{Finley2017a}). 
We also introduce new methodology to interpret the observations in the context of galactic wind radiative transfer models and thereby constrain properties of the phenomenon.  Throughout this work, we adopt the cosmological parameters reported by \citet{Planck-Collaboration:2016eq}.

\section{Observations} \label{sec:obs}
We observed \tkrs\ ($z = 0.6942$) in a pilot program to study the resolved emission and absorption of \mgtwo\ and \fetwo\ of galaxies hosting known outflows originally analyzed in \citet{Rubin:2014fk}. The SFR-$M_*$\ distribution of the \citet{Rubin:2014fk} sample is shown in Figure \ref{fig.sfr_Mstar}, with those objects showing outflow signatures indicated.
\tkrs\ lies in the upper range of the SFR distribution of the galaxy population.  

\begin{figure*}
\centering
\includegraphics[width=0.9\linewidth]{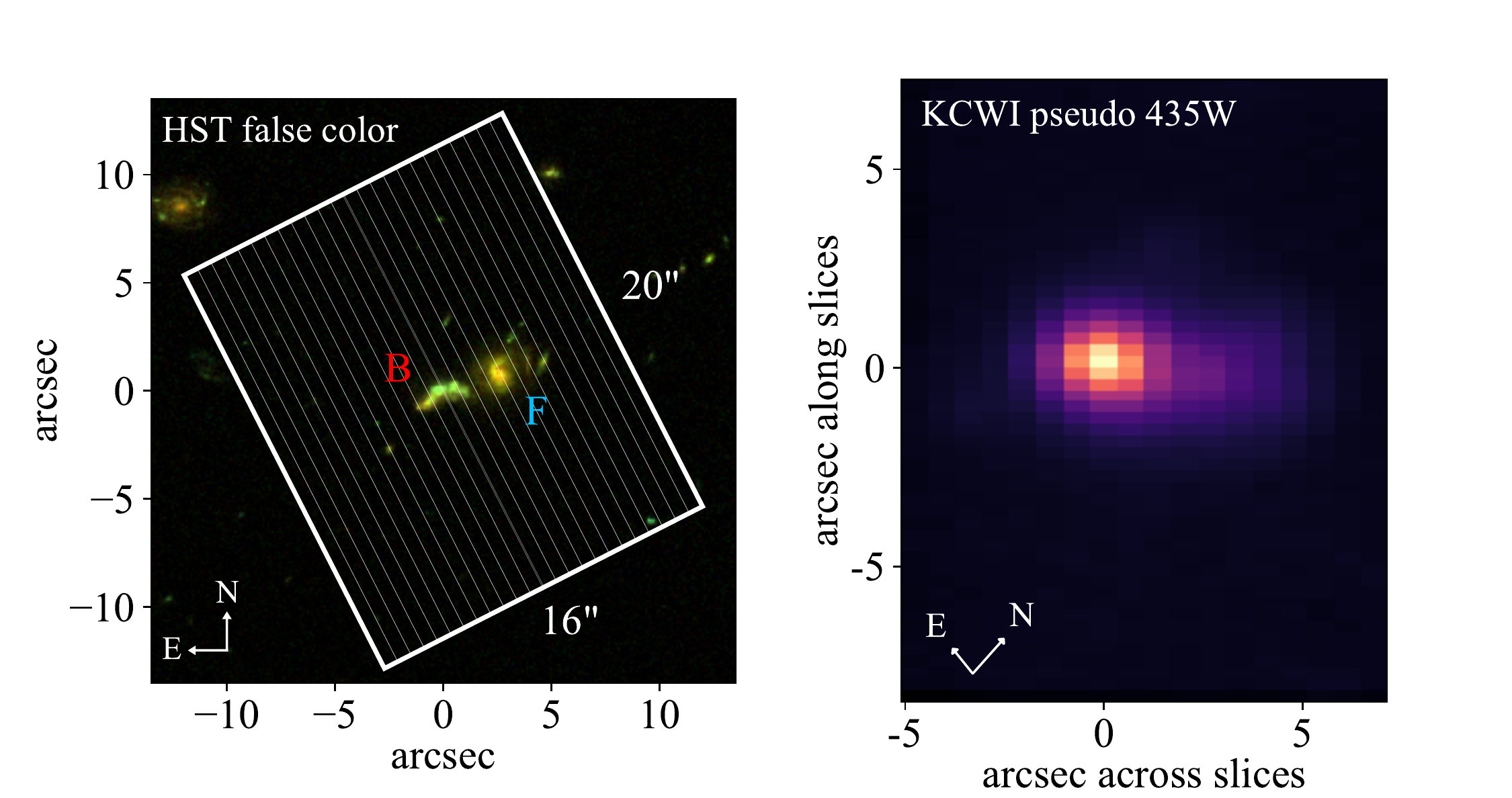}
\caption{\textit{Left:} The primary KCWI slicer position and orientation superimposed on a false-color {\it HST} image composed of the ACS F435W, F606W, and F775W bands.  The two brightest objects in the KCWI field of view are \tkrs, the focus of this paper ($z=0.6942$; labeled `B') and the foreground TKRS 4259 ($z=0.4729$; labeled `F').  \textit{Right:} A pseudo-broadband image from the KCWI data cube FOV using the HST ACS F435W filter response function.  The image has been trimmed to show only the overlapping regions from the dither pattern. Coordinates are shown relative to the brightest pixel at the center of \tkrs. \label{fig.hst_kcwifootprint}}
\end{figure*}

To map the spatially-extended line emission originally discovered by \citet{Rubin2011}, we employed the Keck Cosmic Web Imager integral field spectrograph \citep{KCWI} on the Keck II telescope.  We conducted the observations on the night of 2018 Jan 17 UT under $\sim1\arcsec.4$ (FWHM) seeing conditions as measured at the beginning of the night\footnote{Note: our adopted seeing FWHM = $1\arcsec.6$ as derived later in this section.}.  Our observational setup included the M slicer, providing a $16\arcsec \times 20\arcsec$ field of view (FOV) and sampling of $0.70\arcsec$/pix (width of slices). Combined with the BL grating, this yields a spectral resolution $\mathcal{R}\sim 1800$ (FWHM).  At the redshift of \tkrs, the FOV covers approximately 117 $\times$ 146 kpc along the \tkrs\ major and minor axes, respectively.  We oriented the slicer such that the $0.7\arcsec$-wide slices were oriented along the minor axis (PA = 27$^\circ$) to yield $0.3\arcsec$/pix sampling along the length of each slice; note that this yields rectangular 0.7\arcsec $\times$ 0.3\arcsec pixels and hence pixels will not appear square in images herein.    Figure \ref{fig.hst_kcwifootprint} shows the slicer orientation relative to a high-resolution, false color image of \tkrs\ and other galaxies in the field. 
Table \ref{tab:observ} summarizes information regarding our observations. 

\begin{table}
\caption{Target and observations}
\label{tab:observ}
\begin{tabular}{cc}
\hline \hline
Target name & TKRS4389 \\
Coordinates (J2000) & 12:36:19.84, +62:12:52.9 \\
$z$ & 0.6942 \\
Date of observation & 17 Jan 2018 UT \\
Telescope/Instrument & Keck II/KCWI \\
Slicer/grating & Medium/BL \\
Exposures & 7 $\times$ 1800s  \\
\hline 
\end{tabular}
\end{table}

We obtained 7 $\times$ 1800s exposures on source, dithering with $1.5\arcsec$ offsets between each exposure.  Reduced data cubes were extracted according to standard procedures within the KCWI Data Reduction Pipeline (kderp\footnote{https://github.com/Keck-DataReductionPipelines/KcwiDRP}) except for the sky subtraction step (Stage 5), where we employed a custom sky subtraction algorithm optimized for this field.  Stages 6-8 of the pipeline (including flux calibration) were implemented in the standard manner.

We then corrected the reduced datacubes for known astrometric errors in the WCS solution by creating white-light images for each cube, i.e.,\ summing the flux along the spectral dimension over $\lambda = 4000-5500$ \AA.  The spaxel in each intensity and variance cube corresponding to the brightest pixel of its white-light image was assigned the equatorial coordinates of \tkrs.  We then aligned the cubes by projecting each into a larger cube whose central spaxel was placed at the coordinates of \tkrs, adopting a simple nearest-neighbor shift for partial spaxel offsets.  The WCS-corrected, aligned intensity cubes were then coadded using a weighted mean.   The weights are calculated as follows: we construct a white-light image by summing all spaxels over all wavelengths except for 200 \AA\ buffer regions on the red and blue ends, identify as `bright' pixels those with flux greater than the median flux, sum fluxes and variances in these pixels, and calculate the S/N of the summed pixels.  This S/N serves as the weight for each exposure.  Our fully reduced, aligned, and coadded datacube has a median (over the field of view) 1$\sigma$ limiting surface brightness of \surfbrightlim\ measured in a 5 \AA-wide narrowband image.

To obtain a precise estimate of the seeing and assess the extent of line emission, we employed the {\it Hubble Space Telescope} Advanced Camera for Surveys ({\it HST}/ACS) imaging from the GOODS-N field \citep{Giavalisco2004} taken with the F435W filter.  To perform a final WCS correction, we cut out the $30\arcsec \times 30\arcsec$ portion of the ACS F435W image centered on \tkrs, convolved it with a small Gaussian kernel (FWHM = $0.1\arcsec$), and projected the ACS image into the coordinate system of the coadded KCWI data cube using the \textsc{reproject}\footnote{https://reproject.readthedocs.io/en/stable/} Python package.  This yielded the model reference image/WCS. We then extracted a pseudo-broadband image from the KCWI cube using the F435W filter response function, and set the reference CRPIX in the KCWI WCS to its brightest pixel and the CRVAL to the coordinates of the brightest pixel from the model reference image/WCS.  Lastly, for our seeing measurement, we iteratively convolved the original F435W image with a 2D Gaussian with varying FWHM and reprojected each convolution into the \textit{new} KCWI coordinate frame (informed by the  model reference image/WCS).  After reprojection, we once again set the brightest pixels to have the same coordinates (the vastly different pixel scale between ACS and KCWI induces a small offset) and output the reprojected image with its modified WCS.  Least squares minimization of the difference between this output and the coadded datacube yielded a FWHM~$\sim$ \seeing.

Herein, we present several pseudo-narrowband images, which are generally produced by summing the flux over the spectral direction of the datacube between the two wavelengths indicated.  With the exception of the pseudo-F435W image described above (shown in the right panel of Figure \ref{fig.hst_kcwifootprint}), we assume a flat response as a function of wavelength.  A residual background gradient with flux on the order of a few percent of the \tkrs\ signal is present in these narrowband images, and we subtract this residual structure by fitting a two-dimensional, first-degree polynomial to regions outside the detected emission from \tkrs.

\begin{figure*}[ht]
\includegraphics[width=0.25\textwidth]{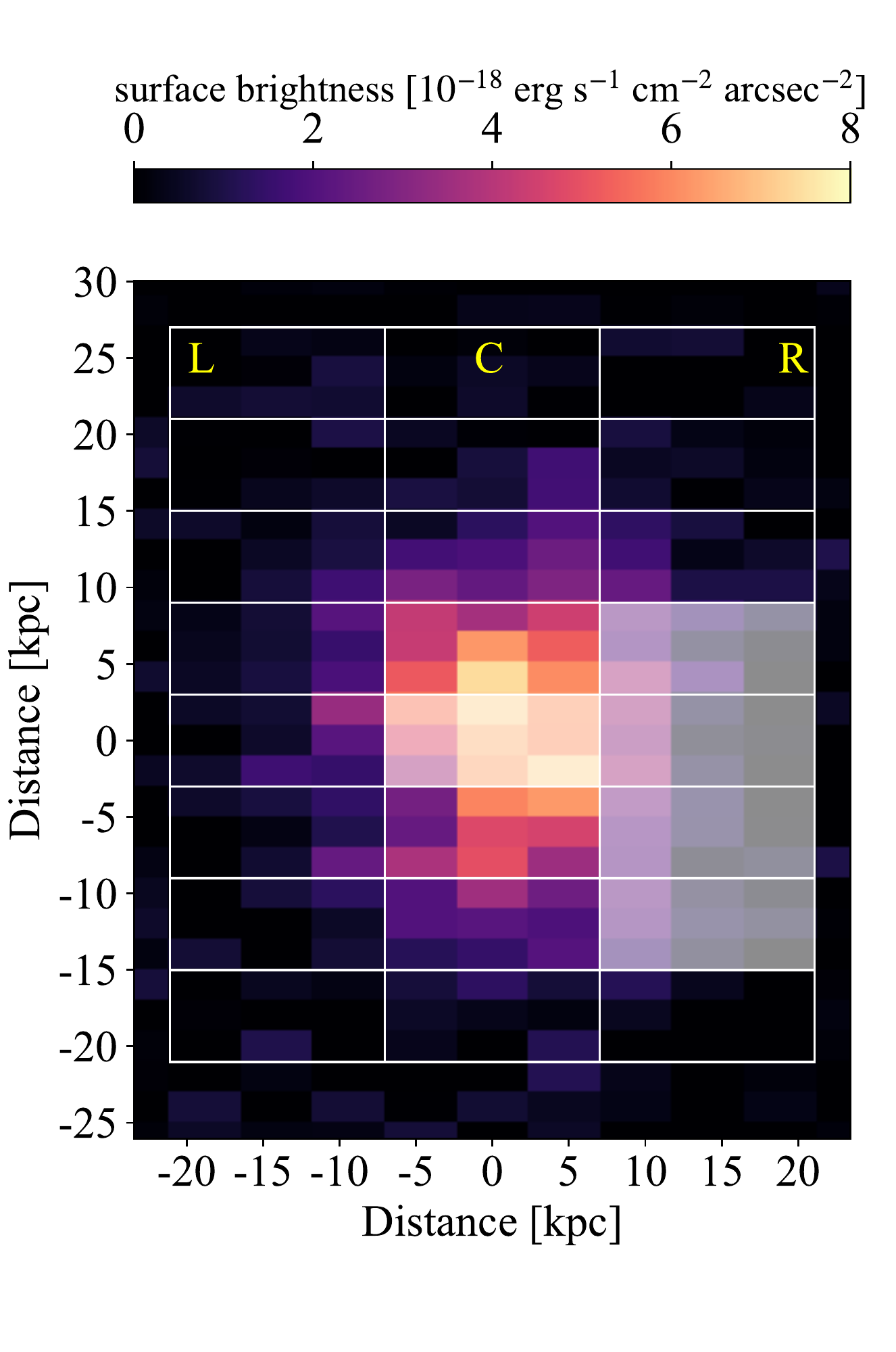}\hfill
\includegraphics[width=0.25\textwidth]{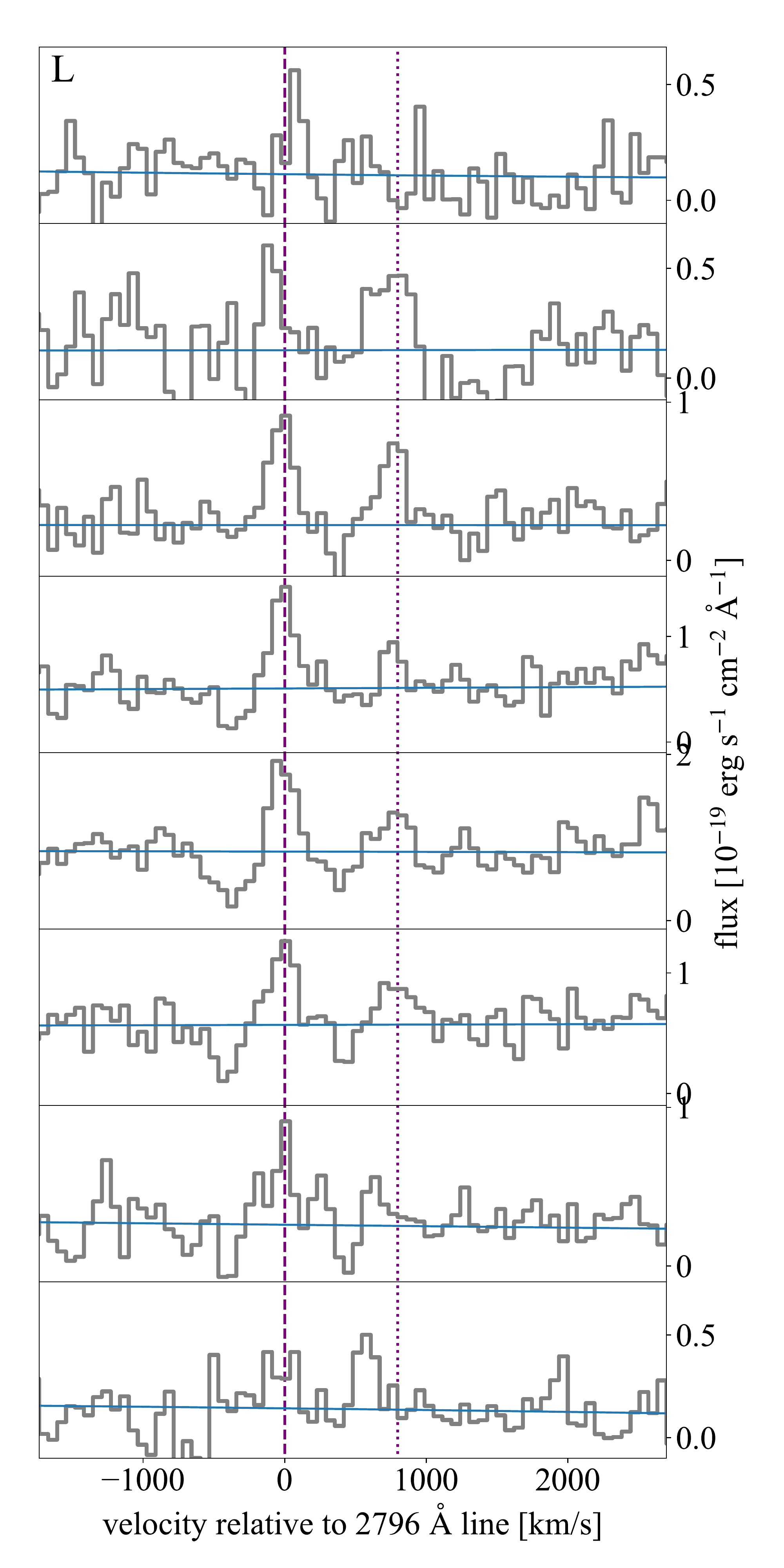}\hfill
\includegraphics[width=0.25\textwidth]{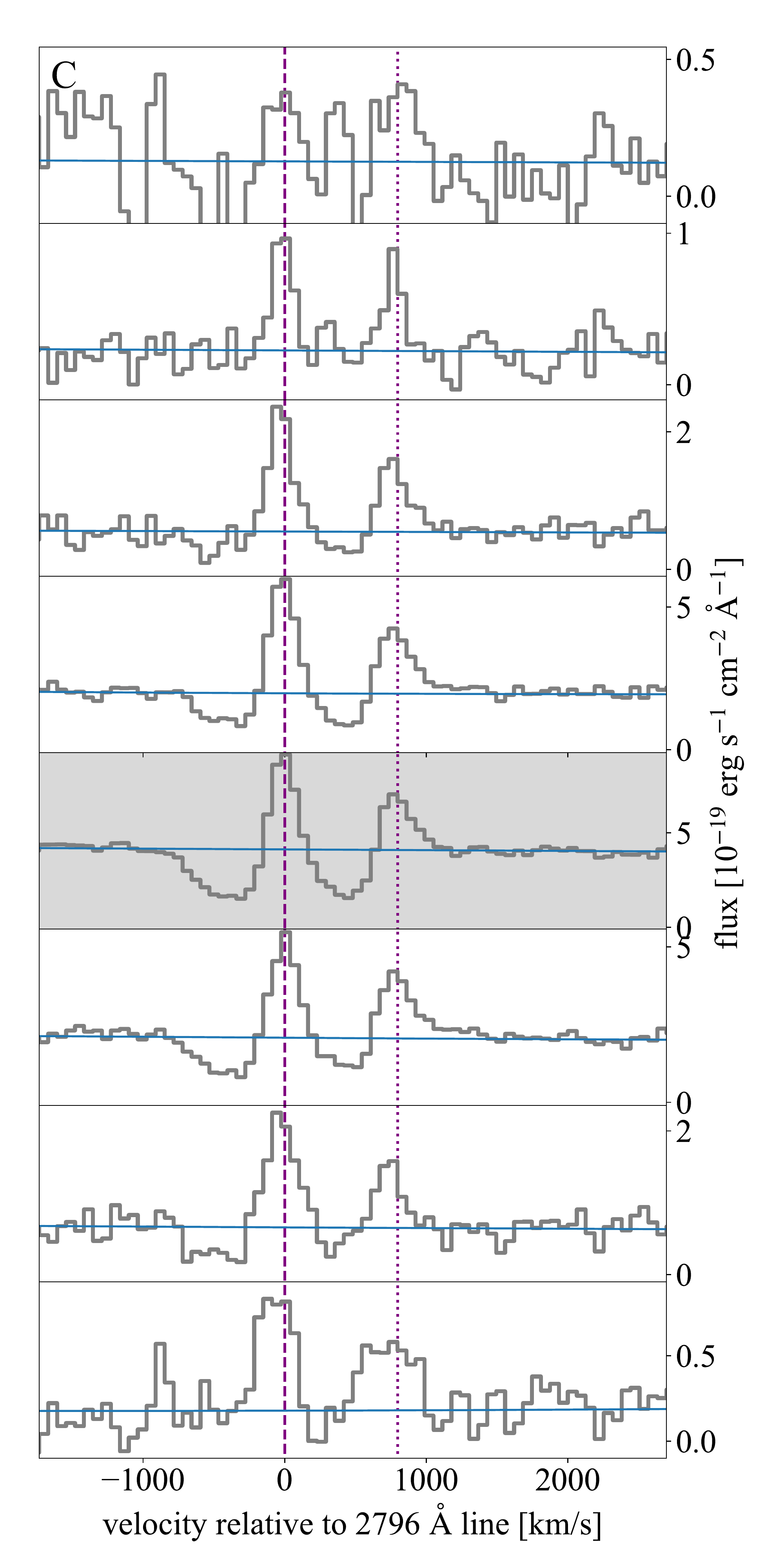}\hfill
\includegraphics[width=0.25\textwidth]{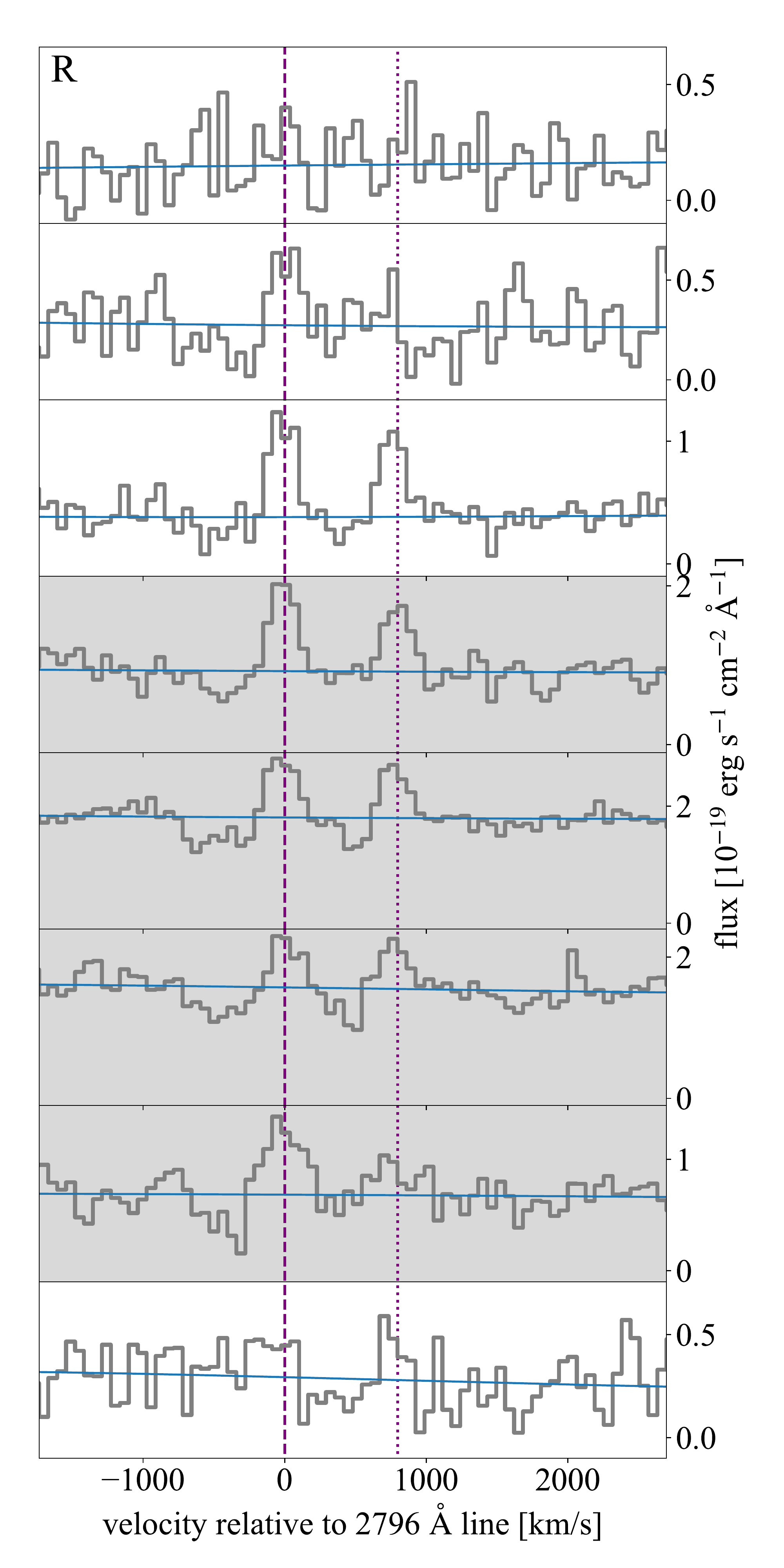}\hfill
\caption{\mgtwo\ emission and absorption from \tkrs\ and its CGM.  The emission map on the left was constructed from a 5 \AA\ slice (approximately 530 $\mkms$) in the spectral direction centered on the $\lambda$ 2796\AA\ emission peak.  Rectangles denote regions over which spaxels were coadded to produce the spectral profiles at right. Corresponding columns in the map and panels with spectra are marked `L', `C', and `R'.  In the three spectra panels, the vertical purple dashed line marks the reference frame set by $\lambda$ 2796 \AA\ at the redshift of \tkrs\ ($z = 0.6942$) and the vertical dotted line denotes the corresponding $\lambda$ 2803
\AA\ location.  Continuua fitted to the spectra using our automated algorithm are shown in blue.  Spectral regions that are grayed out contain spaxels likely affected by the foreground galaxy TKRS4259 (see Section \ref{sec:fittingProcedure}), elevating the continuum level.  We observe significant emission in regions distributed at nearly all azimuthal angles relative to the galaxy's minor axis, suggestive of an isotropic outflow. \label{fig:mapSpectra}}
\end{figure*}


\section{Analysis} \label{sec:analysis}




\subsection{Extended \mgtwo\ emission from \tkrs}
\label{sec:extendedEmission}
From Figure~\ref{fig.hst_kcwifootprint}, we find that
both the {\it HST} image (left) and KCWI pseudo-435W (right) images of the \tkrs\ field exhibit stellar continuum emission from \tkrs\ itself ($z = 0.6942$) and a foreground galaxy TKRS4259 ($z=0.4729$).  A subsequent publication will analyze the spatially-resolved absorption profile of the foreground galaxy CGM \citep[see][for an in-depth study of this TKRS4259]{Rubin:2010_GPG}, as we focus here on \tkrs.  \citet{Rubin2011} showed that not only does \tkrs\ exhibit line emission and absorption from \mgtwo\ in a ``P Cygni'' profile cospatial with the galactic disk, but that the line emission also extends beyond the regions cospatial with the stellar continuum.  These prior observations, conducted via long-slit spectroscopy, were only able to measure emission along the (spatial) slit direction.  With KCWI integral field spectroscopy, we obtain an array of spaxels enabling spectral analysis within each `pixel' of a 2d image.  

We first search for and quantify the extent of the \mgtwo\ emission reported by \citet{Rubin2011}.  To remove contributions from the stellar continua of \tkrs\ and TKRS4259 (which overlaps with \tkrs\ at the spatial resolution of our KCWI data), we construct a continuum-subtracted datacube wherein the spectral direction is sliced to contain a 200\AA\ region (4650-4850 \AA) roughly centered on the observed wavelength of the \mgtwo\ doublet at the redshift of \tkrs.  We then fitted a continuum in each spaxel over this wavelength range using a series of low-order Legendre polynomials\footnote{As described in Section \ref{sec:obs}, a residual background gradient remains in the data after reduction.  This continuum fitting process partially removes the ambient background signal; i.e., a small `continuum' is present in each spaxel.}, using an iterative sigma-clipping scheme to avoid overfitting absorption or emission, and produced two versions: with the continuum subtracted and with the flux normalized by the continuum. 

The left panel of Figure \ref{fig:mapSpectra} shows a 5-\AA\ narrowband image centered on the $\lambda$ 2796 emission peak and extracted from the continuum-subtracted cube.  In the right three panels of Figure \ref{fig:mapSpectra}, we present spectra extracted from 24 regions containing 21 spaxels each ($3 \times 7$) extending to $>20$ kpc above and below the disk plane (with extraction regions marked in the leftmost map).  The velocity scale in each spectral panel is expressed in the reference frame of the \mgtwo\ $\lambda 2976$~\AA\ line.  \mgtwo\ emission from both lines of the doublet is clearly detected in the large majority of these extraction regions.  The physical scale labeled on the axes of Figure \ref{fig:mapSpectra} (left) assumes our adopted cosmology and is expressed relative to the central, brightest pixel of \tkrs\ derived from its stellar continuum image.  

Figure \ref{fig:mapSpectra} shows \mgtwo\ line emission extending well beyond the 7 kpc originally reported by \citet{Rubin2011}.  We now quantify the extent of \mgtwo\ emission detected by our KCWI observations.  In addition to the continuum-subtracted narrowband image described above, we sum the corresponding pixels spectrally in the coadded variance cube in quadrature (over the same 5-\AA\ window) and calculate the detection significance in each spaxel as follows:
\begin{equation}
S^{2796}_{j,k} = \frac{\sum\limits_{i} f_{i,j,k}}{( \sum\limits_{i}~\sigma_{i,j,k}^2 )^{1/2}}
\end{equation}
where $S^{2796}_{j,k}$ is the significance of $\lambda$ 2796 emission in spaxel $(j,k)$, $f_{i,j,k}$ and $\sigma_{i,j,k}^2 $ are the flux and variance, respectively, of the $i$th spectral voxel in the $(j,k)$ spaxel, and the summations run over 5 \AA\ regions centered on the emission peak.      

\begin{figure*}
\includegraphics[width=\textwidth]{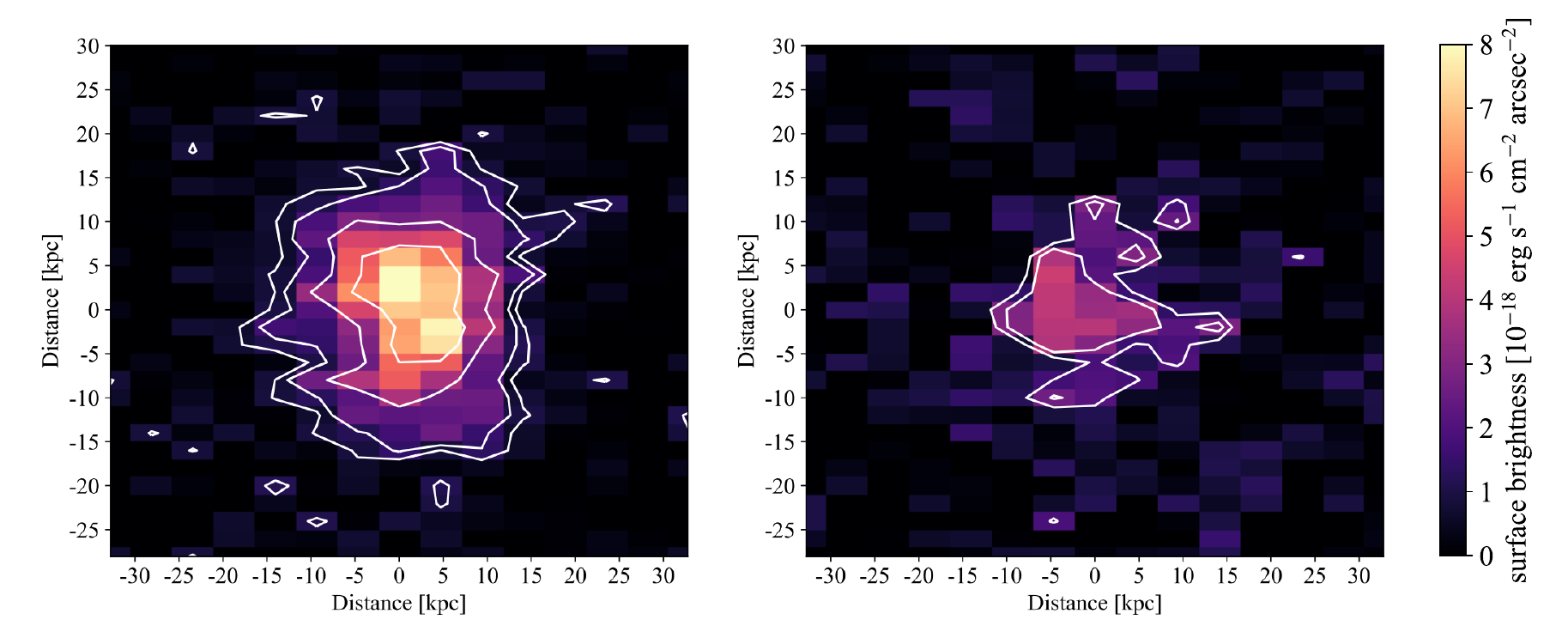}
\caption{Surface brightness of the \mgtwo\ and \fetwostar\  emission extracted from the continuum-subtracted KCWI datacube. Values result from summing the emission over 5 \AA\ and 10 \AA\ windows centered on the $\lambda$ 2796 and $\lambda$ 2626 emission peaks, respectively.  Contours denote 2-, 3-, 6-, and 10-$\sigma$ level detections of the emission (2- and 3-$\sigma$ for \fetwostar).  Using 3$\sigma$ as our detection threshold, the \mgtwo\ emission extends over 37 kpc in the left-hand image. \label{fig:signifEmission} }
\end{figure*}

This calculation results in a significance map, reflected in Figure \ref{fig:signifEmission}, with contours corresponding to 2-, 3-, 6-, and 10-$\sigma$ detection significance.  Adopting a 3-$\sigma$ detection threshold, we measure an extent across the significantly-detected region of $5.0\arcsec \pm 0.4\arcsec$, corresponding to $37 \pm 3$ kpc at the redshift of \tkrs.  This measurement is effectively the angular separation between extreme points in the contiguous 3-$\sigma$ detection region, and we calculate the uncertainty as the half-height and half-width summed in quadrature.  No systematic errors related to, e.g., the WCS offset procedure described in Section \ref{sec:obs} have been included. Below in Section \ref{sec:radprof}, we correct for seeing by deconvolving the emission profile.

\subsection{Fe II* emission}
Our KCWI cube also covers the \fetwo/\fetwostar\ $\lambda$ 2586, 2600, 2612, 2626, 2632 \AA\ multiplet.   In a similar manner to the \mgtwo\ emission, we measured the spatial distribution of continuum-subtracted emission arising from the non-resonant \fetwostar\ 2626 \AA\ transition, chosen because it is intrinsically strongest of the \fetwostar\ lines and is cleanly separated from the resonant \fetwo\ transitions.  Here, we fitted the continuum over 4250-4550 \AA, once again sigma-clipping to avoid overfitting emission/absorption.  A surface brightness/significance map analogous to that of \mgtwo\ is shown in Figure \ref{fig:signifEmission} (right), summed over a 10 \AA\ wavelength window centered on the $\lambda$ 2626 line.  As shown by the contours, we only detect \fetwostar\ emission at 3-$\sigma$ within approximately 7 kpc of the galaxy center (i.e.,\ $\pm 1''$).


\subsection{Radial Profiles}
\label{sec:radprof}
\begin{figure}
\includegraphics[width=1.1\columnwidth]{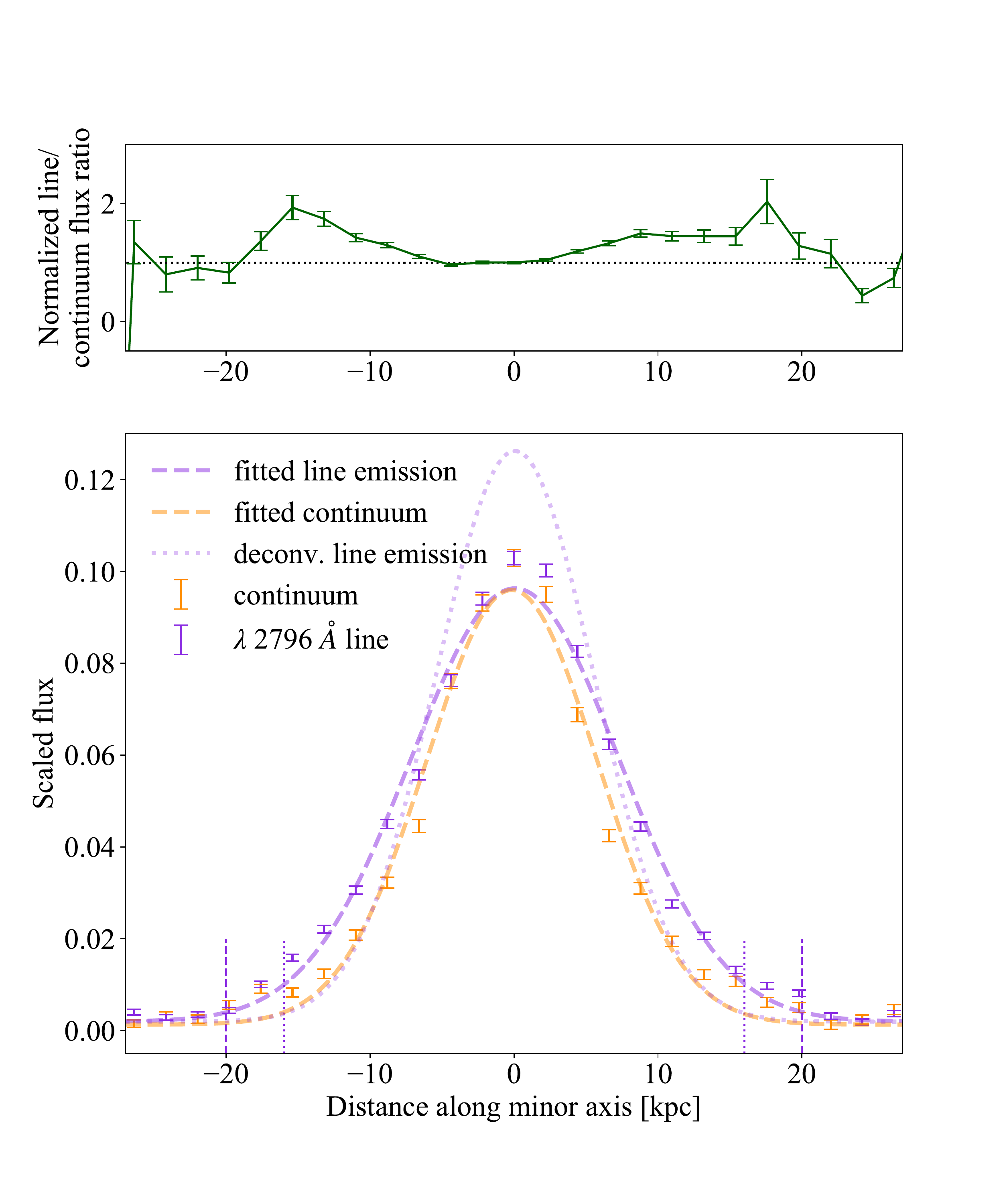}
\caption{{\textit Bottom:} Flux profiles of emission from the \mgtwo\ 2796~\AA\ line (violet) and stellar continuum (orange) above and below the minor axis of \tkrs. The profiles have been scaled to equal amplitude at their peaks. Dashed curves represent Gaussian fits to the line and continuum data. The dotted curve shows the line profile deconvolved with the 1".6 FWHM seeing.  Vertical dashed and dotted lines represent the extent of line emission detected at 5-$\sigma$ in both the measured and deconvolved profiles, respectively.  From the deconvolved profile, we measure an extent of 31 kpc for the \lam 2796 \mgtwo\ emission. {\textit Top:} Ratio of scaled line to continuum profiles along the minor axis.  Both panels show that the line emission significantly exceeds emission from the seeing-convolved stellar continuum and the seeing profile to $>15$ kpc both above and below the major axis. 
The measured line emission at $\approx 10-20$\,kpc offsets 
shows excess emission relative to the stellar component by factors of 30-100\%.
\label{fig:minorAxisProfile}  }
\end{figure}

As an additional measure of the extent of the \mgtwo\ emission, we directly compare the line emission to the stellar continuum in Figure \ref{fig:minorAxisProfile}.  We extracted 5-\AA\ wide pseudo-narrowband images from both the flux and variance cubes (non-continuum subtracted) centered on 1) the \lam\ 2796 \AA\ emission feature and 2) a line-free continuum region approximately 40 \AA\ redward of the \mgtwo\ emission/absorption.  We then summed the flux in 7-pixel wide (by 1-pixel tall) horizontal regions parallel to the major axis and extracted at successive steps along the minor axis, extending to $\sim5.4\arcsec$ above and below the galaxy. The variance cube was similarly summed but in quadrature.  The resulting profiles from the line and continuum emission centered on \tkrs\ are shown in the bottom panel of Figure \ref{fig:minorAxisProfile} with violet and orange points/errorbars, respectively.  Here, we have scaled the peaks of both emission profiles to one another, revealing the excess line emission extending into the circumgalactic regions, to approximately 20 kpc above and below the galaxy.  The top panel plots the ratio of the two flux profiles, showing an increase in the line emission relative to the continuum moving away from the central galaxy to $\gtrsim15$ kpc toward either direction.

From our modeling in Section \ref{sec:obs}, we estimated 1''.6 FWHM seeing.  To control for this potentially significant effect on the measured extent of the emission, we deconvolved the emission profiles shown in Figure \ref{fig:minorAxisProfile} with the seeing as follows: we fitted each with a Gaussian profile (dashed curves), subtracted in quadrature the standard deviation of a 1".6 FWHM Gaussian ($\sigma$=0''.68) from the standard deviation of the fitted profiles, and produced a new Gaussian profile with this new standard deviation.  The resulting deconvolved line emission profile is shown with a dotted curve in Figure \ref{fig:minorAxisProfile}.  We estimated the \mgtwo\ emission extent without seeing effects by locating the outer points with detected emission at 5-$\sigma$ confidence in the original profile and the points in the deconvolved profile with those same flux values.  As a result, we estimate that the emission would extend over $\sim$31 kpc without the effects of seeing.

%
%
%
%
%
%
%
%
%

\section{Comparison to Radiative Transfer Models}
\label{sec:RTmodels}

In principle, the surface brightness profile and relative velocity of the resonant \ion{Mg}{2} emission from this galaxy depends 
on two distributions: 
  (i) the spatial distribution of the source(s); and
  (ii) the
density and velocity of \ion{Mg}{2} ions and distribution of dust in and 
around the object,
which absorb and scatter these photons. We may therefore constrain these latter quantities by comparing the observed emission to that predicted by models that incorporate radiative transfer effects.  An implementation of Monte Carlo radiative transfer in the context of simplified galactic wind models was first described in \citetalias{Prochaska11_RT}.  We use the same code and technique to expand on the original suite of models presented in that work, tailoring the parameter space we explore (in terms of, e.g., wind opening angle and density and velocity laws) to be suitable for our galaxy target.

\subsection{Radiative Transfer Model Suite}\label{subsec.rtmodelsuite}

The wind models of \citetalias{Prochaska11_RT} assume a spherical continuum source with a flat spectrum (i.e., with continuum flux density $f_{\lambda} \propto \lambda^0$).  The radius of this source is assumed to be smaller than the inner radius of the gas composing the wind. \citetalias{Prochaska11_RT} adopted a ``fiducial'' wind model with power law density and velocity profiles:

\begin{eqnarray}
    n_\mathrm{H}(r) = n_\mathrm{H}^0 \left(\frac{r_\mathrm{inner}}{r} \right)^{2} \\
    \vec{v} = v_r (r)\hat{r} = \frac{v_0 r}{r_\mathrm{outer}} \hat{r},
\end{eqnarray}

\noindent where $r_\mathrm{inner}$ is the aforementioned inner radius (set to 1 kpc), $r_\mathrm{outer}$ is the maximum extent of the wind (set to 20 kpc), $n_\mathrm{H}^0 = 0.1~\rm cm^{-3}$ is the hydrogen density at $r_\mathrm{inner}$, and $v_0 = 1000\mkms$ is the wind velocity at $r_\mathrm{outer}$.  The wind is assumed to have a velocity dispersion dominated by turbulent motions with a Doppler parameter $b_\mathrm{D} = 15\mkms$.
A wind metallicity of $Z = 0.5 Z_{\odot}$ is adopted, with dust depleting the Mg by a factor of 1/10.  

As noted in \citetalias{Prochaska11_RT}, these choices 
were made in part so that the resulting model spectra would be similar in shape to \mgtwo\ profiles commonly observed in $z\sim0.5-1$ galaxy spectroscopy \citep{Weiner2009,Rubin2011}, 
but they do not have special physical significance.  Indeed, numerous other models are explored in \citetalias{Prochaska11_RT}, including those with larger values of $n_\mathrm{H}^0$, different density power laws (e.g., $n_\mathrm{H}(r) \propto r^{-3}, r,$ and $r^2$), and different velocity profiles (e.g., $v_r (r) \propto r^{-2}, r^{-1},$ and $r^{0.5}$).  
The parameter space for these models is vast, and here we do not attempt a complete exploration of this space due to the computational expense of the radiative transfer simulations and of preparing the modeled spectra for comparison to our dataset (see \S\ref{subsec:comparison}).  

\begin{figure*}[htb]
\vskip -0.4in
\hskip 0.3in
\begin{minipage}{0.35\textwidth}
\includegraphics[angle=0,width=1.92in,trim=0 0 10 0,clip=]{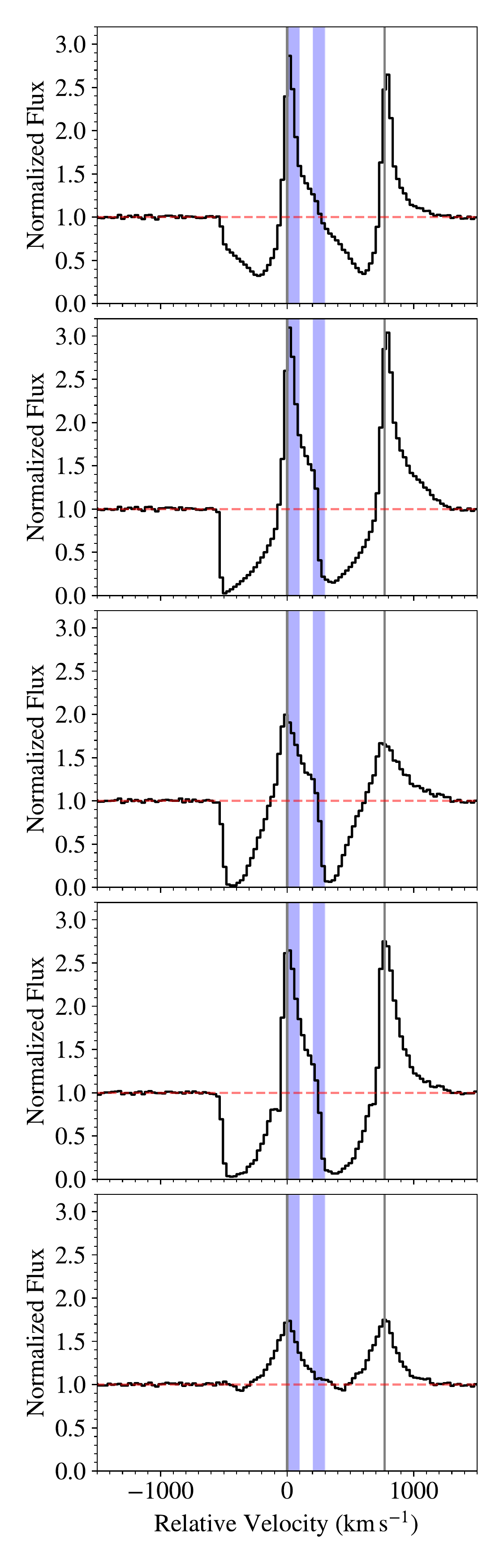}
\end{minipage}
\hskip -0.6in
\begin{minipage}{0.65\textwidth}
\vskip -0.35in
\includegraphics[angle=0,width=5.35in,trim=145 0 0 0,clip=]{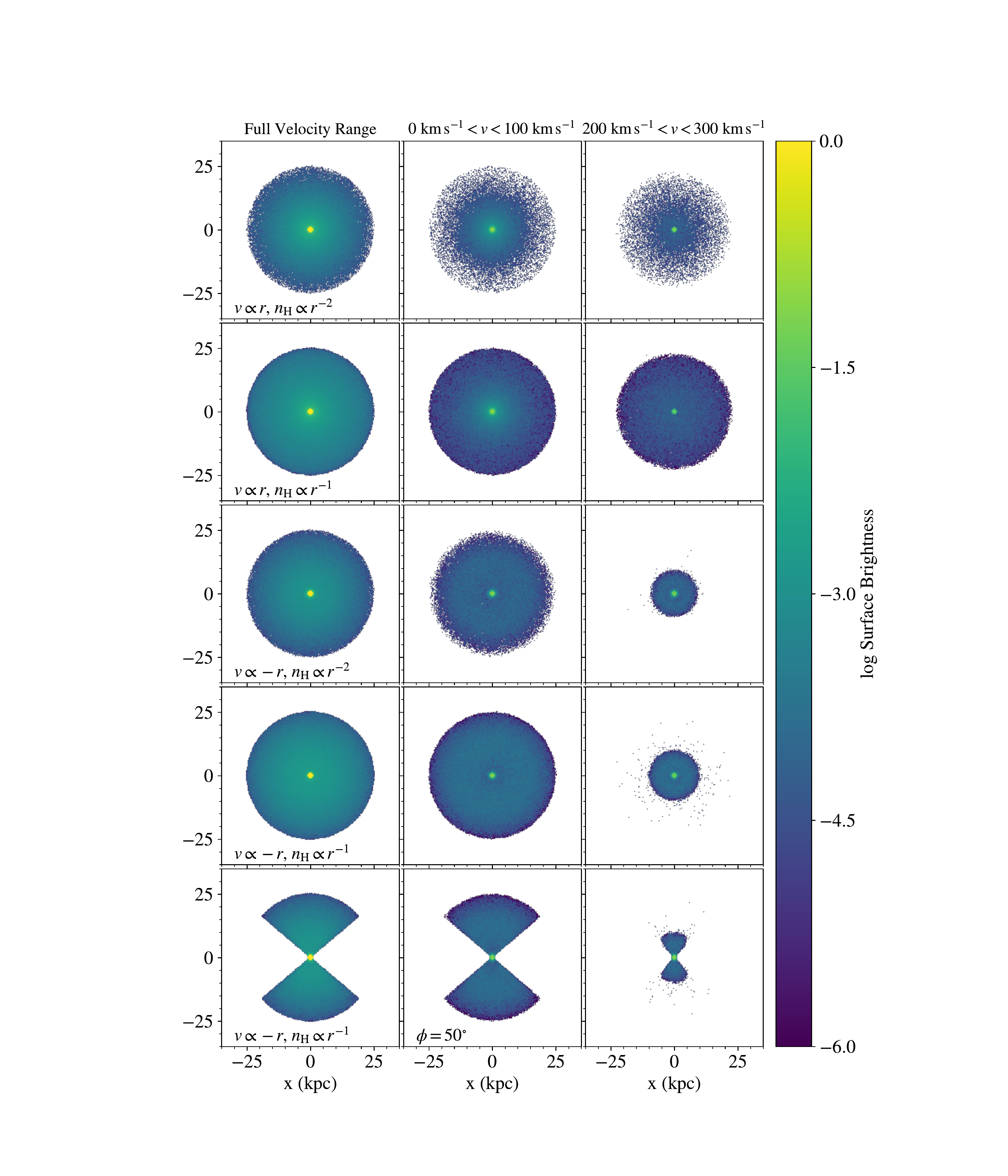}
\end{minipage}
\vskip -0.35in
\caption{The top four rows show spectra and surface brightness profiles predicted for isotropic wind models with the four combinations of velocity and density laws described in \S\ref{subsec.rtmodelsuite}.  The bottom row shows these same properties for a biconical wind model having a half-opening angle $\phi = 50^{\circ}$, and with a decreasing velocity law ($v(r) \propto -r$) and a density power law $n_{\rm H} (r) \propto r^{-1}$.
The spectra shown at left include all photons within the computational domain at wavelengths corresponding to velocities $-1500\mkms < \delta v < 1500\mkms$ relative to the 2796.35 \AA\ transition.  The vertical blue bars indicate the limits of the velocity ranges shown in the images to the right (as labeled on
top).  The vertical gray lines mark the rest velocities of the two doublet transitions.  
\label{fig.rt_demo}}
\end{figure*}

Instead, we began by identifying model spectra from \citetalias{Prochaska11_RT} that reproduce salient features of our observed \mgtwo\ profiles in a qualitative way.  Isotropic models with density profiles scaling as both  $\propto r^{-2}$ and $\propto r^{0}$ tend to yield absorption troughs with apparent optical depths that decrease with increasing velocity offsets (as observed) for several of the velocity laws explored in \citetalias{Prochaska11_RT} (although this is not true in general, as can be seen in Figure~\ref{fig.rt_demo}).  For simplicity, and because it is mass-conserving, we chose to make use of the density scaling $n_\mathrm{H}(r) \propto r^{-2}$ for our model suite.  We adopted a velocity law that increases linearly from $v_{\rm inner} = 50\mkms$ to $v_{\rm outer} = 500\mkms$ from $r_{\rm inner} = 1$ kpc to $r_{\rm outer}$, respectively.  The velocity $v_{\rm outer} = 500\mkms$ is approximately the minimum required to reproduce the observed absorption velocities (after smoothing by the KCWI spectral resolution).  Because a primary focus of our work is on constraining the spatial extent of the wind in TKRS4389, we generated six versions of this model, changing the value of $r_{\rm outer}$ to range between 5 and 30 kpc in increments of 5 kpc.  We adopt solar metallicity for the wind, and assume that dust depletes Mg by a factor $1/10$.

Our initial comparisons of these models to the data (see \S\ref{subsubsec.results}) revealed that they failed to produce emission as strong as that observed at large distances from the central source.  We therefore generated a supplemental set of six models with a shallower density law $n_\mathrm{H}(r) \propto r^{-1}$.  We likewise noted that all of these models tend to yield quite broad, low-amplitude emission lines at large spatial offsets from the continuum source, whereas the emission lines observed in our datacube are narrow in velocity space (with $\rm FWHM\approx 180\mkms$).  In an attempt to simultaneously reproduce both the high velocities observed in absorption and the narrow emission features observed at distances $r \gtrsim 15$ kpc, we generated another set of twelve models, all of which have the same set of density power laws and $r_{\rm outer}$ values as above, but which have a linearly {\it decreasing} velocity law (with $v_{\rm inner} = 500\mkms$ and $v_{\rm outer} = 50\mkms$).  

Finally, motivated by the observational evidence that such winds are often bipolar or biconical \citep[e.g.,][]{Weiss1999,Bordoloi2011,Kornei2012, Bolatto2013, Rubin2014}, we generated several models that emulate this morphology.  Holding $r_{\rm outer}$ fixed at 25 kpc, for each density and velocity law described above, we modified the gas density distribution to be biconical with half opening angles of $\phi = 10^{\circ}$, $30^{\circ}$, $50^{\circ}$, $70^{\circ}$, and $80^{\circ}$.  This yields an additional set of $5\times 4 = 20$ models, for a total of 44 models when combined with those described above.



Model spectra are calculated using the three-dimensional Monte Carlo radiation transfer code of \citet{Kasen2006} as described in detail in \citetalias{Prochaska11_RT}.  Briefly, the values of the wind density and velocity are mapped onto a three-dimensional Cartesian grid with a pixel size of 0.2 kpc.  Each simulation follows the paths of $N\sim10^7$ photon packets as they are absorbed and re-emitted throughout the grid until they exit the grid region. All photons with wavelengths between 2770 \AA\ and 2830 \AA\ are included in the output, and are binned to a dispersion of 0.25 \AA.
For simplicity, all images and spectra of each model are generated at a viewing angle $\theta = 90^{\circ}$ for consistency with our edge-on view of the galaxy target. This choice is unimportant for our analysis of the isotropic wind models; however, changes in viewing angle can produce qualitatively different output spectra for biconical winds.  For example, all such models produce little to no absorption when viewed at $\theta = 90^{\circ}$, whereas if the outflow cones were tilted such that more of the wind material were aligned with our line of sight to the continuum source, they would produce progressively stronger absorption \citep[e.g.,][]{Carr2018}.  We will explore these nuances in future analyses.

Figure~\ref{fig.rt_demo} displays spectra and surface brightnesses predicted for a subset of these models with $r_{\rm outer} = 25$ kpc.  Total surface brightness maps, along with maps showing the surface brightness predicted for two $100\mkms$-wide wavelength windows are shown.  The primary effect of varying the wind density profile between $n_{\rm H} \propto r^{-2}$ and $r^{-1}$ is to boost the surface brightness in the wind outskirts for the latter models (shown in the second,  fourth, and fifth rows).  As seen by comparing the rightmost panels of Figure \ref{fig.rt_demo}, winds that decline in velocity with radius produce much weaker emission in the wind outskirts at wavelengths corresponding to relative velocities $\delta v > 200\mkms$.

\subsection{Direct Comparison to KCWI Spectroscopy}\label{subsec:comparison}

The outflow geometry, densities, velocities, etc., dictate both the projected surface brightness profile of line emission and the spectral profiles observed across the system. Therefore, to achieve the most direct comparison between the radiative transfer models described above and our KCWI data, we compare both the spatial and spectral dimensions simultaneously.  Among several key differences between the raw model and KCWI cubes, the model outputs have much higher spatial and spectral resolution than the data, even without considering the effects of seeing.  Below, we detail how we prepared the models for comparison as well as the results that followed from this comparison.

\subsubsection{Procedure}
  
\label{sec:fittingProcedure}
A number of operations are required to manipulate the datacube products of our radiative transfer model software so that they may be effectively compared to the KCWI datacube.  First, we convolve the `image' predicted within each 0.25 \AA-wide wavelength bin with a two-dimensional Gaussian kernel with FWHM $= \seeing$ (as derived from {\it HST} imaging; see Section \ref{sec:obs}) to emulate the effects of seeing.  We then rebin each image channel 
to the spatial pixel scale of our KCWI data, approximately 0.68'' $\times$ 0.29'' measured from the WCS in the KCWI data.  The models are expressed in physical distance coordinates (kpc), so we convert to angular distances assuming our adopted cosmology \citep{Planck-Collaboration:2016eq}.  We then add a World Coordinate System (WCS) solution onto the model datacubes with the appropriate KCWI spatial pixel scales.  The reference pixel coordinates were determined by locating the brightest pixel in a 75 \AA-wide (4700-4775 \AA) pseudo-narrowband image constructed from the model cube and setting the coordinates of that pixel to the coordinates of the brightest pixel in a similar pseudo-narrowband image of the observed data.  

At this point, the spectral dimension of the transformed model cubes remain mismatched to the KCWI data and are expressed in the rest frame.  We thus redshift the full wavelength scale to that of \tkrs.  Then, for each spaxel, we convolve the spectrum with a Gaussian line spread function concordant with the spectral resolution documented for the BL grating/medium slicer combination ($\mathcal{R}\sim 1800$).  As a final step, we rebin each spaxel to the same dispersion as the KCWI data (1 \AA/pix).

With the spatially and spectrally convolved and rebinned model cubes now aligned with the KCWI data, we directly compared the models to the data on a spaxel-by-spaxel basis.  For both the model and KCWI cubes, we extracted spectra in 2-spaxel, 0.7" x 0.6" apertures (1 horizontal spaxel and 2 vertical spaxels).  Then, to compare the two, we found scaling factors between the continua in extracted spectra from the models and from the data, where continuum regions were defined as pixels where $4700 < \lambda < 4720$ \AA, which contain neither absorption nor emission features.  We adopted the ratio of the median continuum flux value in the data to that of the model, omitting pixels with negligible continuum flux, as each extracted spectrum's scaling factor.  We then adopted as each model's global scaling factor (over the entire field of view) the median scaling factor of the extracted spectra.  

However, a portion of the KCWI field of view is `contaminated' by the foreground galaxy TKRS 4259 \citep{Rubin:2010_GPG}, which increases the continuum level relative to the line emission in affected pixels.  Fortunately, TKRS4259 is primarily situated to the west-northwest of \tkrs\ and its light overlaps only slightly with the stellar disk of \tkrs\ at the spatial resolution of our data, leaving most of the projected area around \tkrs\ unaffected.  Nevertheless, we identified spaxels potentially contaminated by TKRS4259 by assuming its stellar continuum will contaminate the same spaxels that also contain [\otwo] line emission at the redshift of TKRS4259 ($z=0.4729$), which we also cover with these KCWI observations.  We omitted these spaxels from both the continuum scaling determination and the model-data comparison, as indicated in Figure \ref{fig:mapSpectra}, where regions with contaminated spaxels are greyed out (although uncontaminated spaxels within each region were still used in the fits).  Lastly, because we are primarily concerned with the \mgtwo\ emission arising from the galactic wind and \mgtwo\ emission is likely also produced in galactic H II regions, we omitted 3 spaxels within $\sim 0.4\arcsec$ of \tkrs, centered on the brightest continuum spaxel, from the comparison.

Finally, we adopted a $root-mean-square$ ($rms$) of residual values as our goodness-of-fit metric for each model $i$:
\begin{equation}
    rms_i = \sqrt{\frac{ \sum_{j,k} (\mathcal{D}_{jk} - (s_i\mathcal{M}_{ijk}) )^2 }{N}}
\end{equation}
where $\mathcal{D}$ and $\mathcal{M}$ denote the KCWI data and radiative transfer models, respectively, $i$ and $j$ subscript individual model cubes and individual extracted spectra from each cube (model or data), respectively, $k$ subscripts individual pixels in each spectrum, $s_i$ is the scaling factor for each model cube found as described above, and $N$ represents the total number of extracted spectra $j$ included in the comparison.  The $rms$ is calculated over the 75 \AA\ spectral region from 4700-4775 \AA, covering the observed wavelength regions containing the \mgtwo\ doublet and $\sim 25$ \AA\ continuum regions on either side.  Spatially, we include spaxels over an area approximately 60 $\times$ 70 kpc centered on the brightest continuum spaxel of \tkrs, omitting those labeled as contaminated.

\subsubsection{Results}\label{subsubsec.results}
\begin{figure*}
\centering
\includegraphics[width=0.95\linewidth]{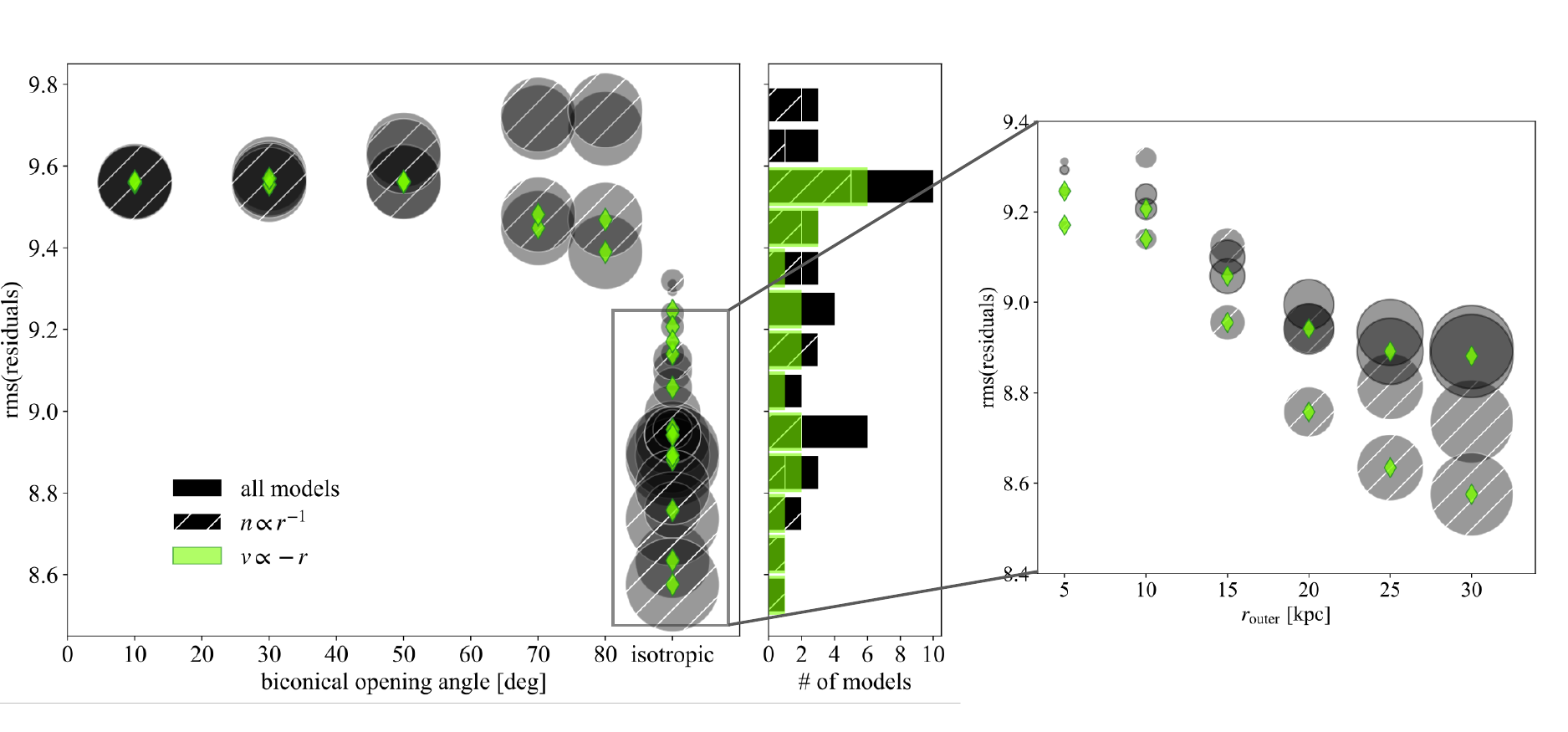}
\caption{Goodness-of-fit for our 3D radiative transfer models given various choices of wind properties.  \emph{Left:} All model $rms$ values are shown as a function of wind opening angle $\phi$ in black circles sized according to $r_{\rm outer}$ (and black histograms in the center panel).  Certain model parameter choices are highlighted such as those with $n \propto r^{-1}$ density profiles rather than $n \propto r^{-2}$, and those with radially decreasing rather than increasing velocity profiles.  \emph{Right:} Showing only the isotropic wind models ($\phi = 90^\circ$), we demonstrate the impact of wind extent ($r_{\rm outer}$), density profile, and velocity profile. The best-fitting models are winds with isotropic geometries, \router$>20$ kpc, $n \propto r^{-1}$, and radially declining velocity profiles. \label{fig:modelsBC} }
\end{figure*}

\begin{figure*}[htp]
\centering
\includegraphics[width=0.33\textwidth]{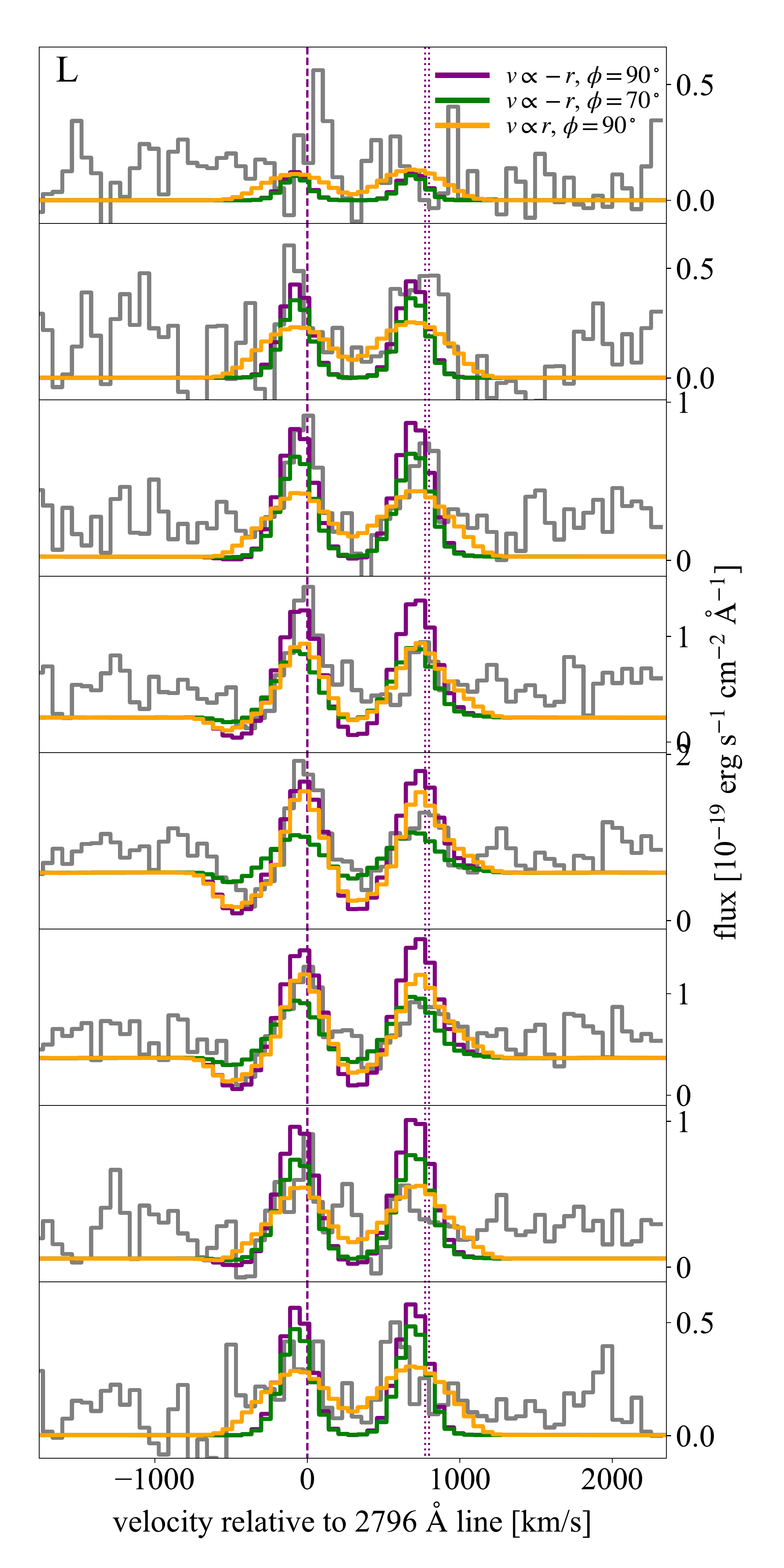}\hfill
\includegraphics[width=0.33\textwidth]{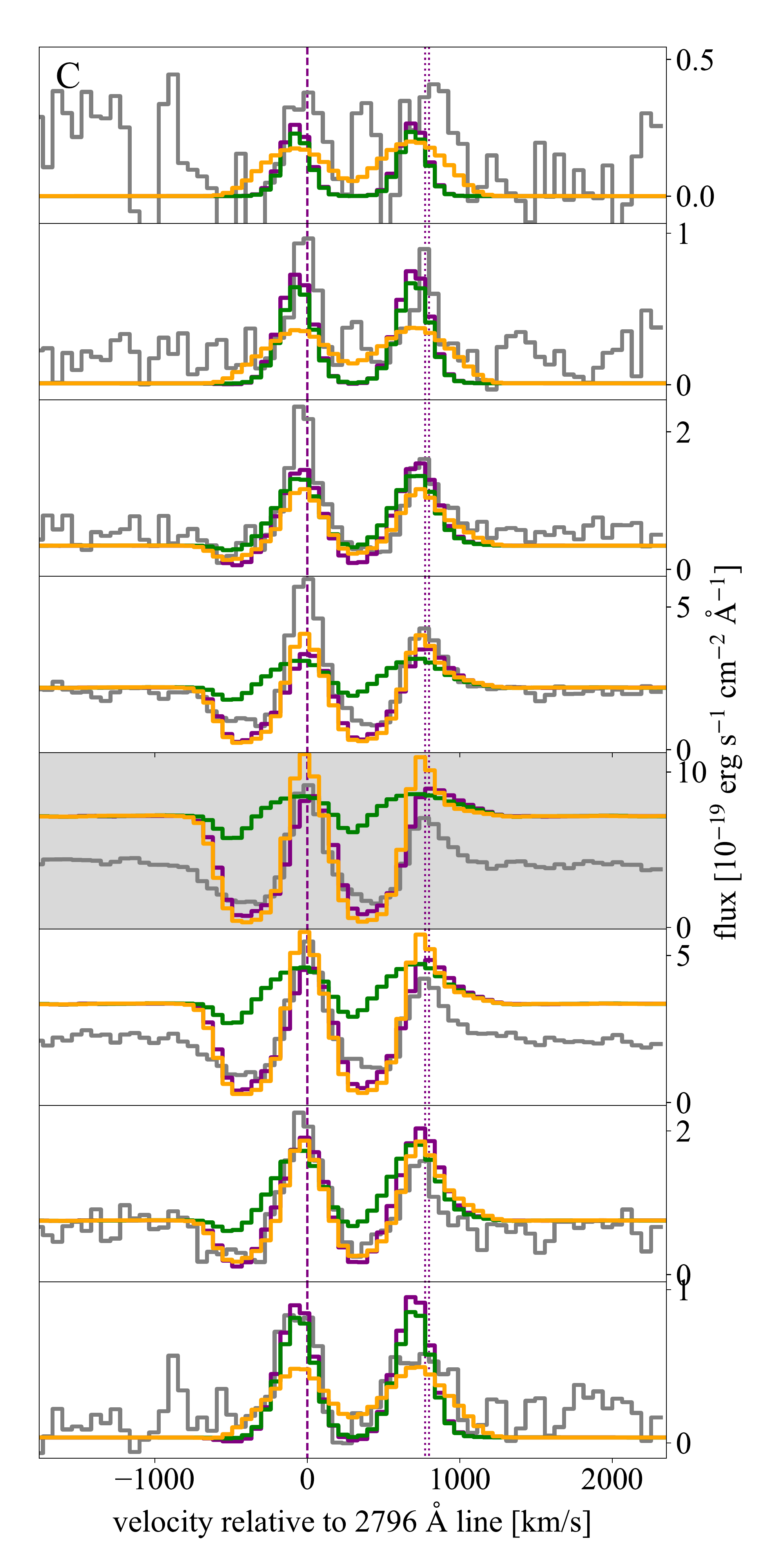}\hfill
\includegraphics[width=0.33\textwidth]{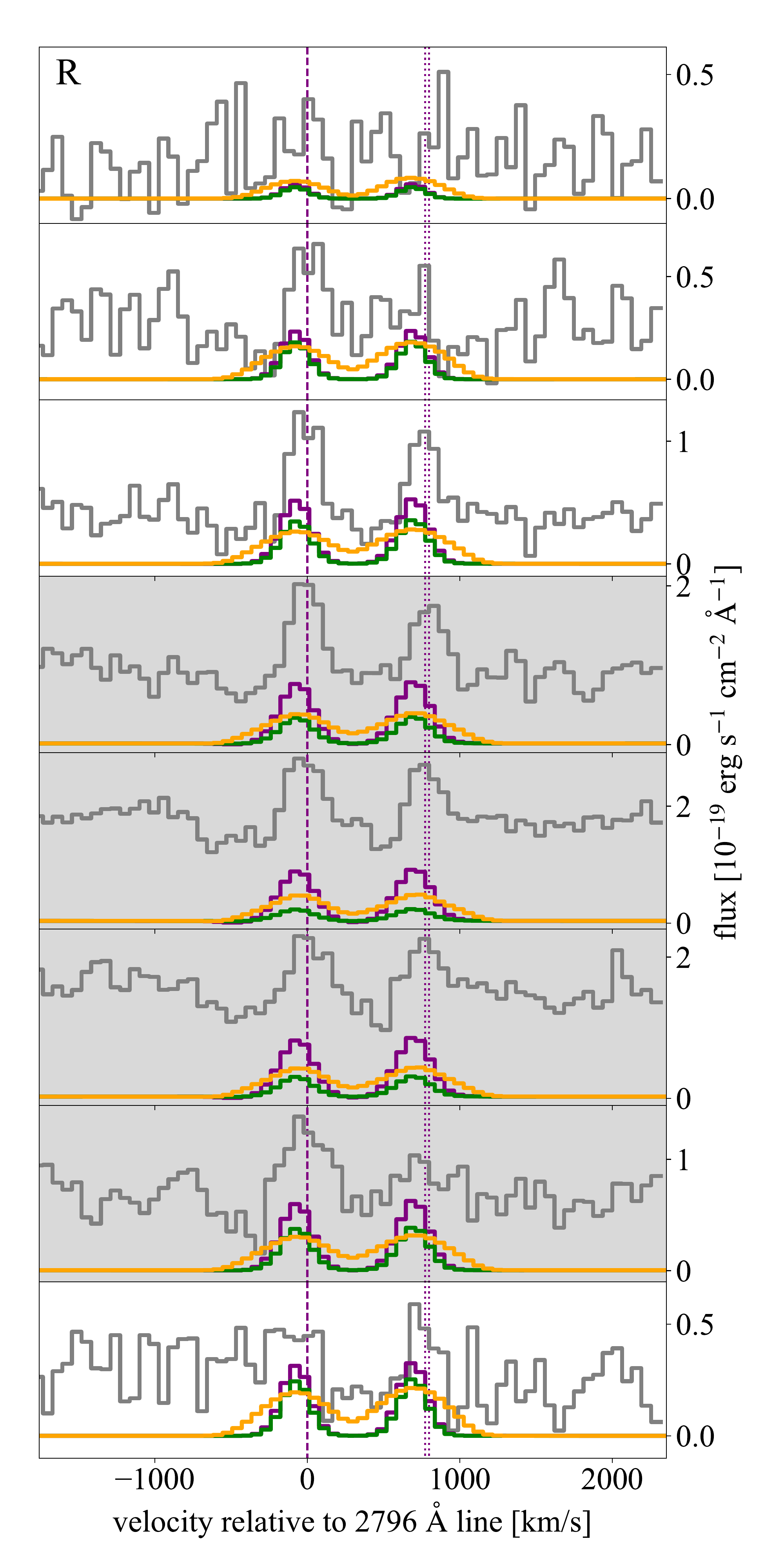}
\caption{Comparison of spectra extracted from our KCWI data with those extracted from our 3D radiative transfer models.  Each column of panels corresponds to a column depicted in Figure \ref{fig:mapSpectra} (left) with the same `L', `C', or `R' label.  Within each panel, we plot three extractions from three models, all of which have $n \propto r^{-1}$ density profiles: (\emph{purple}) the best fit model, which is isotropic and has a radially declining velocity profile and $r_{\rm outer} = 25$ kpc; (\emph{green}) a more collimated wind, also radially declining in velocity and with $r_{\rm outer} = 25$ kpc but with $\phi = 70^\circ$; and (\emph{orange}) an accelerating, isotropic wind with $r_{\rm outer} = 30$ kpc.  This final model is indeed the second-best fitting, and we note that this model poorly matches the data both in the width and height of the emission peaks, especially at large radii.  As in Figure \ref{fig:mapSpectra}, panels containing spaxels omitted from the model fits according to Section \ref{sec:fittingProcedure} have gray backgrounds. \label{fig:modelCompareSpec} }
\end{figure*}

A number of results emerge from our model comparison, summarized in Figure \ref{fig:modelsBC}.  A subset of our convolved models are plotted in Figure \ref{fig:modelCompareSpec}. The left panel of Figure \ref{fig:modelsBC} depicts the 
residual $rms$ values for each of our models with a black circle, with the circle sizes representing the \router\ extent values (larger circles for greater \router).  Particular density ($n \propto r^{-1}$) and velocity ($v \propto -r$) profile shapes are shown with diagonal hatching and green diamonds, respectively.  Histograms of the $rms$ values are shown in the center panel with density/velocity profiles similarly marked, with declining velocities colored green.  The rightmost panel shows only the isotropic models, with $rms$ as a function of wind radial extent.  Most prominently, isotropic models are strongly favored over those with smaller $\phi \leq 80^\circ$; the best fitting is colored purple in Figure \ref{fig:modelCompareSpec}.  Indeed, collimated outflow geometries with $\phi = 10^\circ$ universally show the largest $rms$ relative to the full $rms$ distribution; i.e., changing the extent and velocity or density profiles did not improve the fit because these models produced insufficient flux away from the minor axis.  The isotropic models ($\phi = 90^\circ$) attain the smallest residuals of any opening angle and, indeed, the $rms$ trends downward with increasing opening angle.

Secondly, the modeling provides an additional constraint on the outflow extent to that measured directly in Section \ref{sec:extendedEmission}.  The right-hand panel of Figure \ref{fig:modelsBC} shows only the isotropic models, and demonstrates that the models with the smallest $rms$ values have \router$>20$ kpc.   Those with \router$\leq10$ kpc have insufficient flux at large radii from the galaxy and consistently perform more poorly (regardless of changes in density/velocity profile) with $rms$ greater than the \router$>20$ kpc isotropic models.  Although this result is unsurprising given the results of Section \ref{sec:extendedEmission}, this consistency suggests the extended signal we observe is not the spurious result of, e.g., poorly understood instrumental blurring effects.

Lastly, the models that best fit our data feature \textit{radially declining} velocity profiles (green diamonds in Figure \ref{fig:modelsBC}), with initial velocities $v_{\rm inner} = 500 \mkms$ and final velocities $v_{\rm outer} = 50 \mkms$.  The impact of the wind velocity is most evident in the \mgtwo\ emission profiles at large radii, which we demonstrate in Figure \ref{fig:modelCompareSpec} where we superimpose both data and model extractions from regions illustrated in Figure \ref{fig:mapSpectra}.  In particular, a wind with radially increasing velocity  broadens the emission profile, and the modeled profile exhibits excess flux at velocities offset from the systemic and insufficient flux at the systemic velocity.

\section{Discussion}\label{sec:discussion}
\subsection{Origin of the outflow}
  \citet{Rubin:2010_GPG} and \citet{Rubin2014} report that \tkrs\ exhibits some level of AGN activity, has SFR $\sim 50$ $M_\odot$ yr$^{-1}$, and is undergoing a merger, all phenomena that likely eject gas from galaxies. As described in Section \ref{sec:RTmodels}, our data suggest an outflow that is close to isotropic and strongly disfavor biconical models that are highly collimated.  One may intuitively expect AGN-driven outflows to be relatively collimated due to the minuscule physical scale of a black hole accretion disk relative to its host galaxy and the presence of collimated jets observed at X-ray, optical, and radio wavelengths in some nearby active galaxies.  On the other hand, galaxy-wide star formation episodes naturally provide the spatially-distributed matter and energy injection to drive less collimated outflows via, e.g., supernova, radiation pressure, and cosmic rays \citep[e.g.,][]{Murray2010,Murray2011,Thompson:2015_radPressure}. However, observations of galaxies hosting active nuclei indicate that AGN can indeed drive galaxy-scale outflows \citep{Harrison:2012_outflowsULIRGs, Leung:2017_mosdef}\footnote{However, we qualify that \tkrs\ is very different from these AGN-selected samples, e.g., with an X-ray luminosity $\geq 2$ orders of magnitude fainter \citep{Ptak2007}.}.  \citet{Baron:2018aa} mapped with KCWI an outflow traced by [\textsc{O~iii}] emission from a post-starburst galaxy with an active black hole. Their map reveals a 17 kpc asymmetric conical structure only present to one side of the galaxy, far different to the symmetric wind we observe around \tkrs, and stellar population as well as dynamical modeling indicate an AGN-driven outflow.  Although geometry alone will likely not discriminate between outflow mechanisms, the emerging diversity of wind geometries from spatially-resolved observations will serve as important constraints for both small-scale \citep{Fielding2017} and large-scale \citep{Nelson:2019_outflows} modeling of galactic winds.
  
The irregular morphology and high SFR of \tkrs\ suggest it may share some physical similarities with the class of ultraluminous IR-bright systems (ULIRGs) undergoing concurrent mergers and starbursts at low redshift \citep{SandersMirabel1996}.  While the blue colors of our target at $z\sim0.7$ make it an imperfect analogue of dust-dominated ULIRGs, they may nevertheless have similarities in their wind geometries and driving mechanisms.  The significant observational evidence for the ubiquity of winds traced both in \ion{Na}{1} D and OH absorption and in CO emission in these systems \citep[e.g.,][]{Rupke2005a,Martin2005,Veilleux2013,Cicone2014,Lutz2020} strongly suggests that they likewise have an isotropic geometry (as argued by, e.g., \citealt{Lutz2020}).  The relative importance of star formation vs.\ AGN activity in driving ULIRG winds remains to be firmly established; however, a scenario in which star formation plays an important role in launching winds wherever it is present, with AGN contributing a ``boost" in mass outflow rates that is correlated with AGN luminosity, has emerged in recent years \citep{Veilleux2013,Cicone2014,Fiore2017,Lutz2020}. 
  
 \citet{Nelson:2019_outflows} track outflows driven by both supernovae and AGN feedback in the Illustris TNG50 simulation and find that while both supernova and AGN-driven winds may begin with large, near-isotropic opening angles near the galaxy, the outflows become collimated on larger scales as the wind travels through the CGM.  This effect intensifies at lower redshift in their simulations and is prominent by $z\sim1$; however, the simulated winds are nearly isotropic on the scales we observe here ($\sim20$ kpc) and become more collimated at the scales typically probed in QSO absorption line experiments \citep{Bordoloi2011,Kacprzak2012}.

\subsection{\mgtwo\ Emission Mechanisms}

In our analysis, we have made the assumption that the observed \mgtwo\ emission has arisen purely due to resonant scattering of photons through the cool photoionized outflow driven by the target galaxy.  However, there are at least two other potential mechanisms that may give rise to \mgtwo\ emission from diffuse circumgalactic material: recombination of $\rm Mg^{++}$ ions, and collisional excitation of the \mgtwo\ $\lambda \lambda 2796, 2803$ doublet transitions.  Indeed, collisional excitation is predicted to yield significant emission in other rest-frame UV metal-line transitions tracing the CGM; e.g., \ion{C}{3} $\lambda 977$ and \ion{Si}{3} $\lambda 1207$ \citep{vandeVoort2013,Sravan2016,Corlies2018}.  We therefore assess here the potential contributions of these processes to the observed emission using the spectral synthesis code CLOUDY (version 17.01; \citealt{Ferland2017}).

We consider a simplified scenario of an optically thin, infinite slab of gas having solar metallicity and three different hydrogen volume densities: $n_{\rm H} = 0.1~\rm cm^{-3}$, $0.01~\rm cm^{-3}$, and $0.001~\rm cm^{-3}$.  For our source of ionizing photons, we adopt the extragalactic UV background provided to \citet{Ferland2017} by Haardt \& Madau (2005) (modified from \citealt{HM1996}) and interpolated to $z=0.7$.  For this set of conditions, CLOUDY predicts ionization parameters ranging between $\log U = -2.1$ and $-4.1$
for the lowest and highest volume densities, respectively,
and neutral hydrogen fractions in the range $x_{\rm HI} \approx 10^{-3} - 10^{-5}$.  The resulting slabs range in thickness ($d\ell = N_{\rm HI}/x_{\rm HI} n_{\rm H}$) from $\approx 4$ kpc to $\approx 0.4$ pc, and the predicted \ion{Mg}{2} doublet emissivities range between $10^{-33} - 10^{-27}~\rm erg~cm^{-3}~s^{-1}$.  Assuming these emissivities arise from a $1~\rm kpc^{3}$ cube of gas at $z=0.7$, the resulting \ion{Mg}{2} $\lambda 2796$ surface brightness would be $7.5\times10^{-23}~\rm erg~arcsec^{-2}~cm^{-2}~s^{-1}$ for $n_{\rm H} = 0.001~\rm cm^{-3}$, $1.3\times10^{-20}~\rm erg~arcsec^{-2}~cm^{-2}~s^{-1}$ for $n_{\rm H} = 0.01~\rm cm^{-3}$, and $8.6\times10^{-19}~\rm erg~arcsec^{-2}~cm^{-2}~s^{-1}$ for $n_{\rm H} = 0.1~\rm cm^{-3}$.  The predicted \ion{Mg}{2} $\lambda 2803$ surface brightnesses are $\approx 30-50\%$ lower.  The former two values are well below the surface brightness detection limit of our observations ($1.4 \times 10^{-18}$ erg s$^{-1}$ cm$^{-2}$ arcsec$^{-2}$ at 3$\sigma$), and all are well below the surface brightness of the extended emission we observe at $>2 \times 10^{-18}$ erg s$^{-1}$ cm$^{-2}$ arcsec$^{-2}$.

We caution that the assumption of a 1 kpc-thick cloud is technically inconsistent with our assumption of optically thin conditions, as such a cloud would likely give rise to optically thick \ion{H}{1} at densities $n_{\rm H}\sim0.01-0.1~\rm cm^{-3}$ \citep[e.g.,][]{Rahmati2013}.  A more physically plausible scenario would be one in which there are numerous, small, dense clouds filling a larger-scale hot flow \citep[e.g.,][]{Rauch1999,Crighton2013,Crighton2015,Lau2016,McCourt2018,Gronke2018}.  The surface brightnesses reported above may be considered the maximum values that would arise from this type of cloud geometry.

\subsection{The mass of the flow}

In `down-the-barrel' absorption-line studies of outflows, estimates of mass outflow rates are typically approached as follows \citep{Rupke2005b,Weiner2009,Martin2012,Rubin2014}.  First, a fiducial flow morphology is adopted; often, this is an axisymmetric bicone extending to a distance $r_{\rm outer}$ from each galaxy's center.  The gas is assumed to remain at a constant velocity $v_0$ as it flows outward.  The resulting mass outflow rate may be written
\begin{equation}
    dM/dt = \frac{4\pi}{3}C_{\Omega}\mu m_p N_{\rm H, flow} r_{\rm outer} v_0,
\end{equation}
with $C_{\Omega}$ equal to the angular covering fraction of the bicone, and $\mu m_p$ equal to the mean molecular weight. $N_{\rm H, flow}$ is the total hydrogen column density along a line of sight through the bicone, and is to some extent constrained by the data via analysis of blueshifted metal-line absorption in transitions such as \mgtwo, \ion{Fe}{2}, or \ion{Na}{1}.  However, ionization fractions, metal abundance ratios, and dust depletion factors must  be adopted in order to estimate $N_{\rm H, flow}$ from metal-line column densities, all of which are poorly constrained and uncertain by orders of magnitude.  
The velocity, $v_0$, is the most tightly-constrained quantity in this context, as it may be measured directly from the galaxy spectroscopy.  The extent of the flow, $r_{\rm outer}$, and the morphology of the putative bicone are typically constrained by these data only in the sense that the material must extend over a considerable area of the luminous component of the galaxy in order to give rise to any detectable absorption (and so, e.g., likely extends to at least the half-light radius with an angular covering of $C_{\Omega} \gtrsim 1/2$; \citealt{Weiner2009}; \citetalias{Prochaska11_RT}). 

The emission-line data and analysis presented above eliminate several of the major sources of uncertainty in estimates of the mass flow rate.  First, our analysis directly constrains the morphology and extent of the wind, suggesting it is isotropic (such that $C_{\Omega}=1$) and extends to at least $r_{\rm outer} \ge 20$ kpc.  
For this particular galaxy with a half-light radius $R_e = 3.8$ kpc \citep{Rubin2014}, using these latter values ($r_{\rm outer}=20$ kpc and $C_{\Omega}=1$)  in the above equation increases $dM/dt$ by a factor of ${\sim}10$.  Adopting the \fetwo\ column density measured for the wind in this system by \citet{Rubin2014} of $\log N_{\rm FeII, flow}/{\rm cm^{-2}} > 14.8$, and assuming that all Fe is singly ionized, a solar abundance ratio ($\log \rm Fe/H = -4.49$; \citealt{SavageSembach1996}), and a dust depletion factor of $-1.0$ dex (as in \citealt{Rubin2014}), we find that $\log N_{\rm H, flow}/{\rm cm^{-2}} > 20.3$ \footnote{We choose to make use of our constraints on $N_{\rm FeII, flow}$ rather than $N_{\rm MgII, flow}$ because the absorption line analysis performed in our earlier work did not explicitly model the effects of scattering, which are far more significant in the latter case.}.
Under the assumption of a spherically-symmetric wind extending to $r_{\rm outer}=20$ kpc with $v_0 = -300\mkms$ (and $\mu=1.4$), we estimate $dM/dt \approx 57~ M_{\odot}~\rm yr^{-1}$ -- a value comparable to the SFR of this system.

The simple wind morphology adopted for this estimate, however, is inconsistent with the geometry of the three-dimensional wind model favored by our dataset.  Instead of a wind with a constant density and velocity, the observed \mgtwo\ profile is most closely matched by a wind with a velocity and density that decline as a function of distance.  In particular, the model yielding the smallest $rms$ value as described in \S\ref{subsubsec.results} has velocity and density laws:
\begin{gather*}
    v(r) = -18.75~\mathrm{km~s^{-1}~kpc^{-1}} (r - r_\mathrm{inner}) + v_\mathrm{inner}\\
    n_\mathrm{H}(r) = 0.1~\mathrm{cm^{-3}} (r_\mathrm{inner} / r),
\end{gather*}
with $r_\mathrm{inner} = 1$ kpc and $v_\mathrm{inner} = 500\mkms$ (assuming a solar abundance ratio for Mg, that all Mg is $\rm Mg^+$, and a dust depletion factor of $-1.0$ dex).  Given these profiles, we may calculate the mass flow rate as a function of $r$:  
\begin{equation*}
    \frac{dM}{dt}(r) = 4\pi \mu m_p n_\mathrm{H} r^2  v, 
\end{equation*}
or
\begin{equation*}
    \frac{dM}{dt}(r) \approx 22~ M_{\odot}\mathrm{yr}^{-1} \frac{r}{\mathrm{1~kpc}} \frac{v(r)}{500\mkms}.
\end{equation*}
Due to this diverging density law, the mass flow rate rises from $22~M_{\odot}~\rm yr^{-1}$ at the base of the wind ($r=1$ kpc) to ${\sim}147~M_{\odot}~\rm yr^{-1}$ at $r=10$ kpc, with $v(r=10~\mathrm{kpc})\approx330\mkms$. The flow rate decreases at larger radii, to ${\sim}55~M_{\odot}~\rm yr^{-1}$ at $r=25$ kpc.
The analogous model with an increasing velocity law yields $\frac{dM}{dt}(r=1~\mathrm{kpc}) \approx 2.2~M_{\odot}~\rm yr^{-1}$ and $\frac{dM}{dt}(r=10~\mathrm{kpc}) \approx 97~M_{\odot}~\rm yr^{-1}$.  

For context, the halo mass for this system, estimated using the stellar-to-halo mass relation of \citet{Moster2013} for the stellar mass $\log M_*/M_{\odot} = 9.9$ at $z=0.69$, is $\log M_{h}/M_{\odot} \approx 11.7$.  Such a halo has a virial radius $R_{\rm vir} \approx 143$ kpc (assuming it collapsed at the epoch of observation as is conventional; \citealt{MB2004}).  
Studies of outflows in cosmological zoom simulations such as the FIRE suite \citep{Muratov2015,Muratov2017} have focused on measuring flow rates at distances $\ge 0.25 R_{\rm vir}$, which for this system is beyond the extent of any of our adopted wind models ($0.25 R_{\rm vir} \approx 36$ kpc). They assess the average cumulative mass loading factor $\eta = \frac{dM/dt}{\rm SFR}$ as a function of halo mass and redshift, finding that values of $\eta$ tend to fall in the range $1 < \eta < 10$ for halo masses $M_h \sim 10^{11.7}M_{\odot}$ at $0.5 < z < 2$. This range is consistent with the mass flow rate and resulting value of $\eta \sim 1$ we estimate for distances of $r=25$ kpc above. 

Given our best-fitting wind velocity radial profile, the outflowing wind material is unlikely to escape the gravitational potential of the galaxy, despite the vigorous episode of star formation \tkrs\ is undergoing.  Assuming a $M_h \sim 10^{11.7}M_{\odot}$ halo with a Navarro-Frenk-White profile and the halo concentration parameters from \citet{Ragagnin:2019_haloConc}, we estimate an escape velocity of $195 \mkms$ at 25 kpc.  Thus, the velocity implied by our wind model (50 $\mkms$) falls well short of the escape speed.  Instead, the material is likely to fall back onto the galaxy in a `galactic fountain' \citep[e.g.,][]{Fraternali:2006_fountain,Armillotta:2016uq}.  We add the caveat that our third-best fitting model, with a radially increasing velocity profile, would in fact exceed the escape velocity at this radius.

More broadly, the wide range in the estimates of mass outflow rate listed above illustrates the importance of wind morphologies and velocity profiles in determining the impact of these flows on their host galaxies. It also implies that a more general and complete exploration of the parameter space of wind density distributions and velocity laws  -- if this were to yield a different ``best-fit" wind morphology -- could result in significantly different mass outflow rates.  Here, we simply report our best estimates of the mass flow rate under the assumption of the adopted density and velocity laws, the combination of which successfully reproduce the detailed features of the observed \mgtwo\ profiles.  Regarding the physical origin of our observed wind, our best-fit model suggests a scenario in which either the wind sweeps along material encountered as it leaves the galaxy, or the cool wind mass increases due to in-situ condensation \citep[e.g.,][]{Thompson2016}, as the $n \propto r^{-1}$ profile requires that mass is added as distance increases.  Profiles measured from high-resolution hydrodynamical simulations \citep[e.g.,][]{Schneider:2018_CGOLS} that reproduce this phenomenon could yield helpful comparisons for interpreting observations such as ours.   

Future work must compare the range of plausible wind geometries to this and other similar datasets in greater depth.  Given the significant evidence that the distribution of \ion{Mg}{2}-absorbing material in galactic winds and gaseous halos is  characterized by clumpy, small-scale structures \citep[e.g.,][]{Monier1998,Rauch2002,RogersonHall2012,Chen2014,Crighton2015,Lau2016,Lopez2018}, such explorations should include models that allow for clumpiness in the wind material \citep[e.g.,][]{Schneider:2018b,Hummels2019,Suresh2019}.  \citet{Gronke2017} have performed a detailed study of the Ly$\alpha$ spectral profiles predicted by radiative transfer models through winds having numerous small, optically-thick clumps along the line of sight.  They find that if the number of clumps exceeds a threshold $\gtrsim10-50$, the resulting spectra can be successfully fit with much simpler, homogeneous shell-model geometries, and furthermore that the column density of the best-fit shells are similar to those of the clumps in total.  These results suggest that the simple wind model fitting we adopt here may indeed be yielding columns and mass flow rates that are representative of the physical characteristics of the wind in TKRS4389.  However, any similarity between the line profiles produced by homogeneous vs.\ clumpy winds, as well as any correspondence between the implied column densities of such models, remain to be demonstrated for \mgtwo.  




\subsection{\tkrs\ in Context with Previous Surveys}
\citet{Finley2017b} and \citet{Feltre:2018aa} presented \mgtwo\ and \fetwo\ emission and absorption results from the MUSE Hubble Ultra Deep Field Survey, where they analyzed these line profiles for a sample of galaxies across the star forming main sequence \citep[SFMS;][]{Whitaker2011}.  Intriguingly, \citet{Finley2017b} found that galaxies that exhibit only emission (and no absorption) in \mgtwo\ and those with \fetwostar\ emission occupy separate regions of the SFMS:  galaxies with $M_* < 10^9~M_\odot$ exhibit \mgtwo\ emission but no \fetwostar\ emission, and those with $M_* > 10^{10}~M_\odot$ exhibit \fetwostar\ emission and strong \mgtwo\ absorption but no \mgtwo\ emission. Galaxies with intermediate masses show emission from both \mgtwo\ and \fetwostar\ and tend to have \mgtwo\ P-Cygni profiles. These authors also found that those galaxies with P-Cygni profiles have higher SFR surface densities than those that show pure \mgtwo\ emission. 
\citet{Kornei2013} found in composite spectra that \mgtwo\ emitters tend to have higher specific SFRs, lower masses, and less reddening due to dust, the latter of which would directly impact photon propagation and thus the emission mechanism.  Albeit at higher redshift ($z\sim2$), \cite{Erb:2012aa} found a similar relationship between galaxy stellar mass and the presence of \mgtwo\ emission and absorption. 
\tkrs, with its $M_* = 10^{9.9}~M_\odot$ and 
$\Sigma_{\rm SFR} = 0.956~M_\odot$ yr$^{-1}$ kpc$^{-2}$
\citep{Rubin2014}, falls into the ``intermediate" stellar mass regime identified by \citet{Finley2017b}.  Its strong P-Cygni line profile is typical of such systems, as well as those with comparably high SFR surface densities.


More recently, \citet{Henry2018} studied both Ly$\alpha$ and \mgtwo\ line profiles in a sample of ``Green Pea" galaxies using a combination of {\it HST}/COS and MMT optical spectroscopy.  They found the \mgtwo\ line profiles exhibited strong emission and either weak or negligible absorption.  In principle, such profiles must be dominated by \mgtwo\ line emission from \ion{H}{2} regions, as the total equivalent width of absorption plus emission features resulting from pure continuum scattering must sum to zero (in the absence of slit losses and/or dust scattering). \citet{Henry2018} concluded that the observed line profiles are consistent with a picture in which this \ion{H}{2} region emission is subject to a modest level of resonant scattering due to the low optical depth of the ISM and outflow material in these very low-mass, starbursting systems.

\citet{Guseva:2019_mg2} analyzed a large sample of low-metallicity, star-forming galaxies observed in the Sloan Digital Sky Survey, selected based on the strength of their H$\beta$, [\ion{O}{3}], and \mgtwo\ emission.  
They reported that galaxies exhibiting \mgtwo\ emission lie above the SFMS (have high SFRs relative to their stellar masses).  They attribute this trend to the observed \mgtwo\ photons arising from \textsc{H~ii} regions, where the \mgtwo\ emitters have relatively greater recent starburst activity.  This finding lends support to similar conclusions reached by \citet{Henry2018} and \citet{Feltre:2018aa} 
based on photoionization modeling (see also \citealt{Erb:2012aa}).  While our radiative transfer modeling assumes that the extended emission we observe arises from resonant scattering of \emph{continuum} photons by a galactic wind, these previous analyses suggest that some contribution from \textsc{H~ii} regions is likely in the central galactic regions.  Indeed, some excess \lam\ 2796 \AA\ emission is apparent at small distances ($<10$ kpc) from \tkrs\ relative to our radiative transfer models; this excess emission may originate from \textsc{H~ii} regions hosting vigorous star formation.

We note that the \citet{Finley2017b}, \cite{Feltre:2018aa}, and \citet{Henry2018} studies characterize the \mgtwo\ and \fetwo\ emission in a `down-the-barrel' sense, analyzing spectra dominated by the stellar galactic regions and not by emission from the CGM.  Drawn from the same galaxy sample, \citet{Finley2017a} reported \fetwostar\ emission extending to a half-light radius of $\sim4$ kpc around a galaxy whose disk extends to a half-light radius of only 2.5 kpc.  In Figure \ref{fig:signifEmission} (right), we show \fetwostar\ emission extended $\sim7$ kpc from the center of \tkrs, but caution that our effective resolution is too poor to properly resolve these scales.  Regarding the detection or lack thereof of extended \mgtwo\ in the \citet{Finley2017b} sample, we note that their redshift range extends from $z = 0.85-1.50$ and peaks at $z\sim1.05$.  Therefore, the higher redshifts alone (relative to the redshift of \tkrs\ at $z\sim0.69$) would induce a $>5\times$ attenuation of the emission due to cosmological surface brightness dimming.  

Only a handful of previous studies have directly assessed the spatial extent and surface brightness of \mgtwo\ emission in the regions surrounding galaxies' stellar components. Although they are unable to detect extended emission around individual galaxies, \citet{Erb:2012aa} stacked 33 2D spectra to reveal excess \mgtwo\ line emission over the continuum on $\sim1\arcsec$ scales at the 1.5-2$\sigma$ level. \citet{RickardsVaught:2019} observed five galaxies at $z\sim0.7$ with narrowband imaging using filters covering the wavelength region of the redshifted \mgtwo\ doublet. They did not detect extended \mgtwo\ emission to limits of $\sim 6 \times 10^{-19}$ \sbunits, sufficiently sensitive to detect the \tkrs\ circumgalactic emission.  Three of their four galaxies with high-quality measurements have higher stellar mass than \tkrs, while the fourth has similar or slightly lower stellar mass.  This object, however, has an SFR approximately 1/5 that of \tkrs, and its lack of detection fits a scenario where CGM \mgtwo\ emission may be anticorrelated with galaxy mass (at least above some mass) but correlated with SFR. As discussed above, similar trends are exhibited by the \mgtwo\ coincident with star-forming regions (i.e., not in the CGM).  

A difference in geometry may also have an effect, as \citet{RickardsVaught:2019} suggest that anisotropy may reduce the emission below their detection limits.  Our modeling strongly suggests an isotropic wind around \tkrs, but this is likely not a generic characteristic given the biconical geometries implied by absorption line studies \citep[e.g.,][]{Kacprzak2012}.  Like \citet{Rubin2011}, \citet{Martin2013} used slit spectroscopy to observe extended \mgtwo\ emission around a high-SFR galaxy, reporting spatially asymmetric extended \mgtwo\ to $\sim 12$ kpc.  This galaxy is more similar to \tkrs\ than most of the \citet{RickardsVaught:2019} sample, which fits the stellar mass and SFR dependence picture outlined above.  However, the radically different morphology implies that while one may expect to observe galactic winds via extended \mgtwo\ for galaxies of certain properties (e.g., $M_*$ and SFR), they likely possess diverse geometries.

\section{Conclusions}\label{sec:conclusions}
We have presented the first observations from our integral field spectroscopy  survey of \mgtwo\ emission and absorption from galactic winds in and around star-forming galaxies.  Our target, \tkrs, is a starbursting galaxy merger at $z=0.69$ with a stellar mass log~$M_*/M_\odot = 9.9$ and SFR = 50 $M_*$ yr$^{-1}$.   
Our spatially-resolved spectroscopy reaches a 1-$\sigma$ surface brightness limit of \surfbrightlim\ measured over a $5-$\AA\ wide spectral region, enabling the first map of extended \mgtwo\ emission on scales extending $>10$ kpc from stellar galactic regions.  From our measurements of the spatial distributions of \mgtwo\ and \fetwo* emission, analysis of the spectral line profiles, and 3D radiative transfer modeling, we report the following key results:
\begin{enumerate}
    \item We discover significant (at the 3-$\sigma$ level above our surface brightness limit) \mgtwo\ $\lambda 2796$ emission arising from the circumgalactic medium and extending $\gtrsim 15$ kpc from the galaxy disk.  From the surface brightness contours alone, we measure a total extent of 37 $\pm$ 3 kpc, while we estimate a $\sim31$ kpc total extent along the minor axis after deconvolving with the 1$''$.6 seeing.  
    
    \item In spaxels covering the galaxy disk, where stellar continuum is evident, we observe the characteristic P-Cygni absorption/emission profile of gaseous outflows in \mgtwo, but this feature gives way to full emission in spaxels off the galaxy disk.  The emission extends in all directions from the galaxy, i.e., not simply along the minor axis, suggesting that the outflowing wind is isotropic.    
    
    \item On the galaxy, our data show absorption in the resonant \fetwo\ transitions and emission from non-resonant \fetwostar\ transitions.  However, the \fetwostar\ emission is not significantly extended as is the \mgtwo, having a 3-$\sigma$ extent of $<7$\,kpc from the galaxy and consistent with the derived seeing FWHM.  
    
    \item We have generated a suite of 3D radiative transfer models of galactic outflows and rebinned and convolved the model output both spatially and spectrally to match the KCWI observations.  By fitting each model to the data in three dimensions, we find that isotropic outflow models with radially declining velocity profiles and extents $>20$~kpc are favored.  Among the two density profiles we test ($n\propto r^{-1}$ and $\propto r^{-2}$), the former provides a closer match to the observed emission profile.
    
    \item Our modeling most disfavors biconical winds with small opening angles, those with radially increasing ($v \propto r$) velocity profiles, those with $n \propto r^{-2}$ density profiles, and those with radial extents $r\leq~10$\,kpc.  
    
    \item Independent of our resonantly-scattered outflow models, we considered alternate scenarios in which either collisional excitation and/or radiative recombination give rise to our detected \mgtwo\ signal.  The expected surface brightnesses from these mechanisms fall short of that observed by factors of $\sim2-10^5$ depending on the assumed hydrogen density, further supporting the resonant scattering origin.
    
    \item Our preferred wind geometry, extent, and velocity and density profiles imply a mass outflow rate of $\sim55~M_\odot$ yr$^{-1}$ at $r=25$ kpc.  This mass outflow is approximately equal to the \tkrs\ SFR~$\sim50~M_\odot$ yr$^{-1}$, resulting in a mass loading factor of $\eta \sim 1$ at this radius. We posit that this material is unlikely to escape the galaxy's potential well, as the velocity of our best-fit model wind at $r=25$ kpc (50 km s$^{-1}$) falls well below the escape speed expected for the galaxy halo.  

\end{enumerate}
The observations presented herein represent the first reported integral field spectroscopic map of extended \mgtwo\ emission and the first of a larger survey spanning galaxy stellar mass and SFR.  Thus, these first direct constraints on outflow morphology and extent should inform both subgrid feedback prescriptions employed in cosmological simulations and more detailed theoretical studies of outflow mechanisms themselves.  As the KCWI and MUSE IFUs have become workhorse instruments on 10-m class telescopes, further insights will no doubt follow from similar studies, as resonantly-scattered photons in outflowing winds provide beacons of this fundamental evolutionary activity in galaxies.

\acknowledgments
\section*{Acknowledgements}
The authors wish to recognize and acknowledge the very significant cultural role and reverence that the summit of Maunakea has always had within the indigenous Hawaiian community.  We are most fortunate to have the opportunity to conduct observations from this mountain.  The data presented herein were obtained at the W. M. Keck Observatory, which is operated as a scientific partnership among the California Institute of Technology, the University of California and the National Aeronautics and Space Administration. The Observatory was made possible by the generous financial support of the W. M. Keck Foundation.  We would like to thank Brice M\'enard, Bill Mathews, and David Koo for helpful discussions as well as Zheng Cai for guidance on the data reduction and analysis.

%

\vspace{5mm}
\facilities{Keck(KCWI)}


\software{\\ 
KDERP (https://github.com/Keck-DataReductionPipelines/KcwiDRP) \\
linetools (https://github.com/linetools) \\
astropy \citep{Astropy:2013}\\
reproject (https://reproject.readthedocs.io/en/stable) \\
}

\bibliography{adssample}



\end{document}